\documentclass[twocolumn,preprintnumbers,amsmath,amssymb,aps,pra,longbibliography]{revtex4-1}
\usepackage[T1]{fontenc}
\usepackage{amssymb,amsmath,bm}
\usepackage{graphicx}
\usepackage{color}
\usepackage{bbold}
\usepackage{bbm}
\usepackage{appendix}
\usepackage{soul}
\usepackage[dvipsnames]{xcolor}
\usepackage{ulem}
\usepackage{float}
\usepackage{textcomp}
\usepackage{ mathrsfs }
\usepackage{siunitx}%
\usepackage{braket}

\begin{document}
\title{Topological enhancement of exciton-polariton coherence with non-Hermitian morphing}
\author{Ruiqi Bao}
\email[Corresponding author:~]{ruiqi002@e.ntu.edu.sg}

\author{Huawen Xu}

\author{Wouter Verstraelen}

\author{Timothy C. H. Liew}\email[Corresponding author:~]{timothyliew@ntu.edu.sg}
\affiliation{ Division of Physics and Applied Physics, School of Physical and Mathematical Sciences, Nanyang Technological University, Singapore 637371, Singapore}

%\date{\today} % Leave empty to omit a date

\begin{abstract}
%The non-Hermitian skin effect (NHSE) has been intensely investigated over the past few years and has unveiled new topological phases, which have no counterparts in Hermitian systems. Here in this work, we first develop a scheme to obtain NHSE in a waveguide where all exciton-polariton modes are localized at the boundary by introducing different decay rates for the two spins. We further consider a defect inside the waveguide and investigate the interplay between NHSE and defect mode. Interestingly, we find that the localized defect mode can be morphed into an extended mode inside the waveguide by tuning the non-Hermiticity degree. Further utilizing the driven-dissipative nature of exciton-polariton, we show lasing occurs in the morphing mode. Moreover, because the morphing defect mode is energetically separated from others, the polariton condensate in this state is supposed to have great spatial coherence. Finally, We show the spatial coherence of the morphing mode is much better compared to that of the normal cavity mode.
The non-Hermitian skin effect (NHSE) has been intensely investigated over the past few years and has unveiled new topological phases, which have no counterparts in Hermitian systems. Here we consider the hybridization between the NHSE in an exciton-polariton waveguide and a localized defect mode. By tuning the non-Hermiticity, we find that the resulting ground-state of the system is both spatially extended and energetically separated from other modes in the system. When polariton lasing occurs in the system, we find an enhanced spatial coherence compared to regular waveguides, which is robust in the presence of disorder.
\end{abstract}

\maketitle
\emph{Introduction.} When cavity photons and quantum well excitons are strongly coupled they morph into hybrid particles known as exciton-polaritons~\cite{RevModPhys.82.1489,RevModPhys.85.299}, which exhibit properties from both of their constituents. The photonic component supports a light effective mass, resulting in an energy separation of modes (in finite-sized systems), which further supports spatial coherence when polariton condensation populates states in a narrow energy range. The photonic part further allows dissipation in the system, which makes the system non-Hermitian. To exist, polariton condensates must be supported by a gain in the system. This is typically arranged by a non-resonant laser exciting hot electron-hole pairs that relax in energy feeding the excitonic component of polaritons. While initially requiring cryogenic cooling, polariton condensation is now routinely observed at room temperature~\cite{doi:10.1126/science.1074464,PhysRevLett.98.126405,Kena-Cohen:2010ue,doi:10.1021/acs.nanolett.7b01956,Su:2020wu}.
With advances in etching techniques, it has become common practice to confine polaritons in arbitrary spatial potentials, including waveguides and lattices. The availability of polariton lattices has supported the field of topological polaritons, where the typical objective is to engineer a bandstructure such that it isolates a small number of states in a bandgap using topology to ensure robustness against structural variations and disorder. Su-Schrieffer-Heeger chains~\cite{doi:10.1021/acsphotonics.0c01958,doi:10.1126/sciadv.abf8049,PhysRevLett.129.246801}, Chern insulators~\cite{PhysRevB.91.161413,PhysRevLett.114.116401,Klembt:2018wu}, valley Hall effects~\cite{PhysRevB.96.165432,doi:10.1126/science.abc4975}, spin-valley Hall effects~\cite{PhysRevB.103.L201406}, antichiral edge states~\cite{PhysRevB.99.115423,PhysRevB.106.235310}, and higher-order topological insulators (corner states)~\cite{PhysRevLett.124.063901,doi:10.1126/sciadv.adg4322,Zhang:20} have all been studied in exciton-polariton systems. In many of these geometries, the different options for coupling polariton states, including the spatial overlap of lattice sites and spin-orbit coupling play a crucial role.

Being non-Hermitian systems, polaritons also support a variety of non-Hermitian topological effects, including: exceptional points~\cite{Gao:2015uy,Gao:2018vv,doi:10.1126/sciadv.abj8905}, bound-state-in-continuum modes~\cite{https://doi.org/10.1002/adom.202102386}, non-Hermitian skin effect (NHSE)~\cite{PhysRevLett.125.123902,doi:10.1021/acsphotonics.1c01425,PhysRevB.104.195301,PhysRevB.103.235306}, non-Hermitian corner modes~\cite{PhysRevB.106.L201302}, and end-mode lasing~\cite{PhysRevResearch.2.022051}. In the NHSE, unlike in the Hermitian case, all eigenstates are localized at the boundary of the system, which suppresses backscattering in propagation~\cite{PhysRevLett.121.086803,PhysRevB.99.201103,PhysRevLett.125.186802,Li:2020vi}. A topological invariant known as the winding number is used to explain the non-trivial topology of NHSE. Furthermore, it has been proposed that combining systems with different topologies allows for morphed states~\cite{PhysRevB.103.195414,Wang:2022vc}. This shows that non-Hermitian topology not only provides opportunities for engineering complex energy spectra but also offers a platform to spatially control the eigenstates.

Engineering of the complex energy spectra is potentially useful for the design of polariton lasers, which aim to exploit the coherence of light emitted when a polariton condensate forms. Here it is important to realize that while simple theoretical models often consider a polariton condensate as a single energy state, real polariton condensates have a finite bandwidth. Already in the equilibrium case, a phase with perfect spatial coherence is impossible in a uniform trap in less than three dimensions at finite temperature~\cite{PhysRev.158.383}. Additional fluctuations from the driven-dissipative nature of polaritons are generally believed to make the situation worse~\cite{Fontaine:2022uk}.
This is especially so in materials used for room temperature operation, where fluctuations due to dissipation are stronger and it is necessary to work with short pulses to avoid sample overheating. The highest coherence of polariton lasers requires isolating the lasing mode in energy, which has been achieved in a trapped geometry (and using a sub-wavelength grating to separated different polarized states)~\cite{PhysRevX.6.011026}. While this technique is good for enhancing temporal coherence~\cite{PhysRevLett.129.147402}, the trapping of the condensate necessarily limits spatial coherence to the trap size. In spatially extended systems, it is common for multiple condensates to appear within an energy bandwidth. While spectrally resolving each mode shows a good spatial coherence of each mode~\cite{PhysRevB.80.045317}, an ideal polariton laser would show spatial coherence without such spectral filtering. It is also understood that disorder limits the range of spatial coherence in spatially extended systems~\cite{PhysRevLett.106.176401}(although it could be compensated with active feedback loops~\cite{Topfer:21}). One could hope that the aforementioned topological polariton systems could aid in enhancing spatial coherence by allowing a spatially extended state to exist in a bandgap robust in the presence of disorder. However, the topological properties of Hermitian systems do not manifest specifically in the ground state (which we take as the desired state for polariton lasing) and the non-Hermitian examples considered so far do not operate with real energy bandgaps.

%In this work, we consider a scheme to realize large area spatial coherence based on a defect coupled to a waveguide that presents NHSE. By considering different decay rates for different spins, the system becomes non-Hermitian and all states collapse at the right side boundary. Further plotting the complex eigenenergy spectra under open boundary condition (OBC) and periodic boundary condition (PBC), all energies from OBC are encircled by PBC which is a signature behaviour of NHSE. The defect provides a localized, energetically separated ground state from other continuous states. By tuning the non-Hermicity degree (decay difference), the localized ground state can be morphed into a spatially extended one and finally become localized at right boundary again. We further demonstrated that polaritons can lase in the morphed extended state when taking into account an incoherent pump and nonlinear loss~\cite{PhysRevB.81.235302}. Remarkably, since the lasing state is both spatially extended and energetically separated from other states, we find that the spatial coherence of the morphed extended state is much larger compared to the normal ground state inside the waveguide.
In this work we consider the hybridization of a polariton waveguide supporting the non-Hermitian skin with the modes of an intentional defect (trap). The aim is to morph the ideal properties of the trap, namely the presence of a ground state well-isolated in energy, with the ideal properties of the skin effect, namely the spreading of its wavefunctions over a controllable localization length. We find theoretically that the system can undergo polariton lasing in the ground state. This allows us to predict an enhanced spatial coherence length compared to that of a regular polariton waveguide. As is expected of most topological systems, we also find robustness of the attained spatial coherence in the presence of disorder.

\emph{Model.} \label{sec:model}
\begin{figure*}
\centering
\includegraphics[width=2\columnwidth]{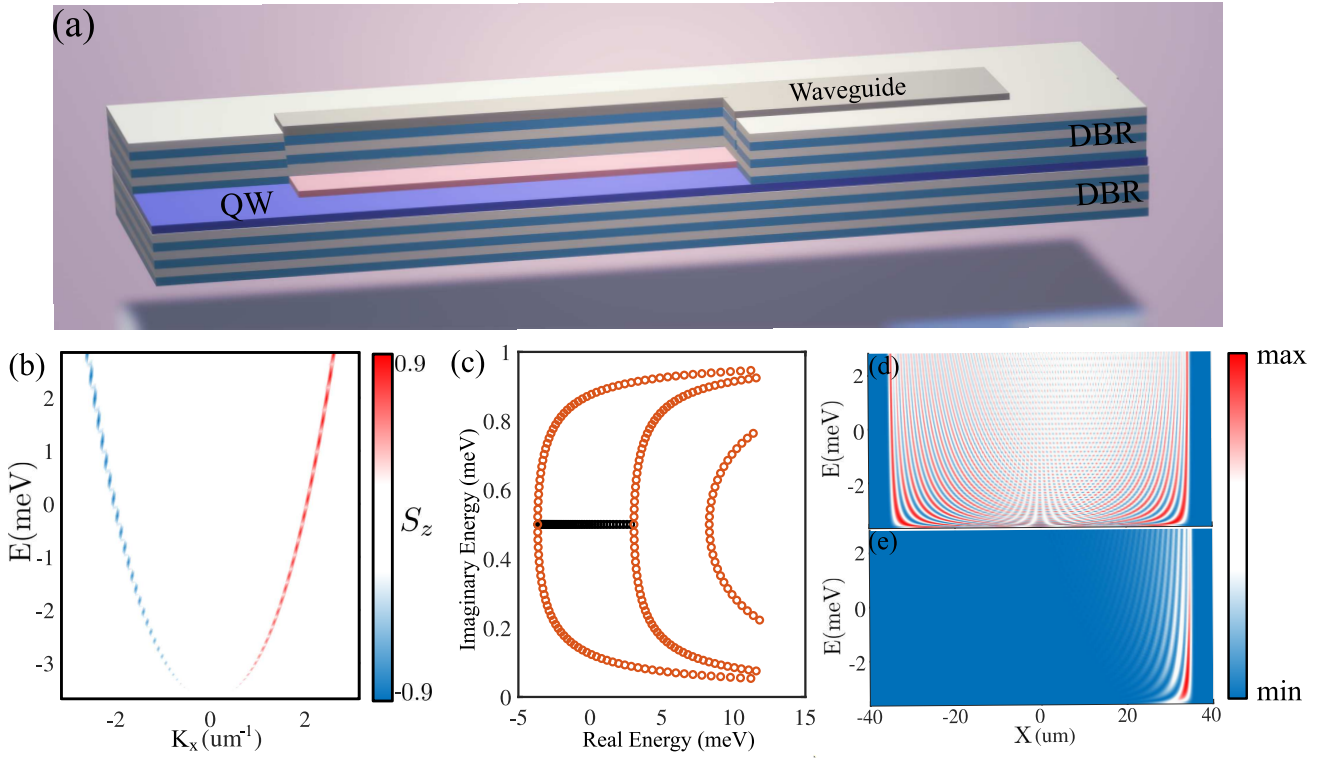}
\caption{(a) The schematic figure of a microcavity with an etched waveguide. The microcavity is formed by two sets of Distributed Bragg Reflectors (DBRs) sandwiching a quantum well (QW). (b) Calculated band structure with the consideration of RDSOC, where $S_z$ represents the circular polarization. (c) Complex energy spectra calculated under different boundary conditions: orange points for PBC and black points for OBC. (d) and (e) Eigenstates' spatial distribution in the Hermitian and non-Hermitian cases respectively. 
Parameters: $m = 2.285 \times 10^{-5}$ of the free electron mass, $\Delta_T = 6~meV$, $\alpha = 1.7~meV$. For (b) and (d), $\gamma_{\pm}=0$; for (c) and (e), $\gamma_+ = 1~meV, \gamma_- = 0$. }
\label{Fig1}
\end{figure*}
It was pointed out in~\cite{doi:10.1021/acsphotonics.1c01425} that the NHSE can be attained in polariton systems by arranging a phase dependent coupling between two chains of modes with different gain/loss. Examples of realizing such a model include using the coupling between spin states in chains of elliptical micropillars~\cite{PhysRevLett.125.123902}, the coupling between vortex modes in chains of polariton rings~\cite{PhysRevB.104.195301}, and most recently the coupling of spin states using Rashba-Dresselhaus spin-orbit coupling (RDSOC)~\cite{PhysRevLett.127.190401,Li:2022uu} in liquid crystal filled microcavities ~\cite{kokhanchik2023nonhermitian}. The latter proposal has pointed out that the NHSE can occur in continuous waveguides, without a lattice. While our results should not depend on the precise underlying mechanism of the NHSE, the latter proposal will serve best our purpose given that spatial coherence was previously considered in similar one-dimensional waveguide ~\cite{PhysRevLett.106.176401,PhysRevLett.113.203902} geometries and it will avoid modulation of the spatial coherence function by the presence of a lattice. Within such a model, the evolution of polaritons is described by:
\begin{equation}
i\hbar \frac{\partial \Psi_{\pm}}{\partial t} = \left( -\frac{\hbar^2 \nabla^2}{2m}+V \mp 2i\alpha\frac{\partial}{\partial x}-i\gamma_{\pm} \right)\Psi_{\pm} + \Delta_T \Psi_{\mp},
\end{equation}

Here $\Psi_\pm$ is the wavefunction of polaritons with $\pm$ spin (corresponding to the circular polarization of emitted light). $m$ is the effective mass; $V$ is a two-dimensional spatial potential profile of our considered microcavity, which is deeper inside the waveguide region ($-20~meV$, grey area as shown in Fig.~\ref{Fig1}(a)); $\alpha$ is the strength of RDSOC; $\Delta_T$ is the X-Y splitting existed in the microcavity; and $\gamma_{\pm} $ correspond to the decay rates of spin up and down polaritons.

%We initially consider the system as Hermitian  by removing the decay term ($\gamma_{\pm}=0$) and obtain the dispersion  by Fourier transforming the eigenstates calculated from Eq.~(1,2). We notice that in the absence of RDSOC, the two spins are degenerate (see supplemental material). After introducing the RDSOC term, the two spins are no longer degenerate due to different spins' dispersion shifting towards different directions in reciprocal space by $2m\alpha/\hbar^2$ ($\Delta_T = 0$). Further adding the X-Y splitting term, we get the dispersion as shown in Fig.~\ref{Fig1}(b). 
It is instructive to first consider the system in a Hermitian regime, neglecting the loss term ($\gamma_\pm=0$), which gives the dispersion shown in Fig. 1(b). The colorbar used in the figure is calculated according to the degree of circular polarization $S_z$, which is defined as: 
\begin{equation}
S_z = \frac{|\Psi_+|^2-|\Psi_-|^2}{|\Psi_+|^2+|\Psi_-|^2}.
\end{equation}
In this situation, time-reversal symmetry is preserved, however, opposite spins propagate in opposite directions. This is a signature behaviour of the topological spin Hall effect, which is further considered in the supplemental material (SM).

Typically in a polariton system, the spin up and spin down decay rates would be the same. However, this changes if we apply a circularly polarized non-resonant pump, which serves as gain. In our calculations, we introduce a spin up polarized, spatially uniform incoherent pump, resulting effectively in $\gamma_+<\gamma_-$. We calculate the eigenstates' spatial distribution and find that, different from the Hermitian case (Fig.~\ref{Fig1}(d)), all eigenstates are localized at the right end of the waveguide as shown in Fig.~\ref{Fig1}(e). Introducing a spin down polarized pump would reverse this localization.
This phenomenon known as the NHSE has no counterpart in Hermitian systems. To prove its topological nature, we calculate the complex energy spectrum under both a periodic boundary condition (PBC) and open boundary condition (OBC) as shown in Fig. 1(c). We find that the complex energy spectrum from the PBC forms a closed loop, which is a signature behaviour of the NHSE~\cite{PhysRevX.8.031079}. In the NHSE, it is expected that all injected polaritons propagate along the same direction (right in this case). The time dynamics of polaritons is further considered in the SM.

%To better understand the NHSE, consider this: as shown in the dispersion, different spins propagate along different directions. Here spin up polaritons, which have a longer lifetime, can survive longer and reach the right end. The left propagating spin down polaritons decay faster, limiting their propagation. 
%In NHSE, backscattering is strongly suppressed which makes it suitable to be used in one way signal propagation.

\emph{Morphing of the defect mode.}
The interplay between the NHSE and the topological edge state in the SSH lattice has been theoretically proposed~\cite{PhysRevB.103.195414} and realized in experiments~\cite{Wang:2022vc}. Here, we consider the interplay between the NHSE and a trivial defect ground state. The defect state is governed by a region with deeper potential than the rest of the waveguide. This could be realized by coupling a micropillar to the waveguide or engineering a region with stronger light-matter coupling~\cite{PhysRevLett.129.147402}.
% has not been studied in the context of polaritons.
%In this work we consider a defect (we take a micropillar as example) which has deeper potential than the waveguide, and study the effect of NHSE on the defect ground state. We want to emphasize that in previous work, the morphing state was obtained by considering non-reciprocal couplings while here the non-Hermiticity is induced by imbalanced decay rates.

Given that the defect has a deeper potential, the ground state of the system could be expected to be localized in the defect. When $\delta \gamma = 0 $ $(\delta \gamma = \gamma_+ -\gamma_-)$ as the case without NHSE, the ground state's spatial profile is shown in Fig. 2(b). The spatial profile of this state changes drastically if we increase $\delta \gamma$. When $\delta \gamma$ is not sufficiently large, the mode will be slightly pulled into the waveguide area as shown in Fig. 2(c). Upon further increasing the decay difference, we find the mode extends within the whole waveguide at the right of the defect (Fig. 2(d)). Eventually, with a larger $\delta \gamma$, the mode becomes localized at the right end, similar to other modes within the waveguide (Fig. 2(e)). 
This effect occurs due to the competition between defect's trapping and localization of the NHSE. The localization length of the NHSE depends on the value of $\delta \gamma$.
The skin modes become more localized when $\delta \gamma$ increases. The enhancement of localization at the right end compensates for the decay of the defect mode resulting in the delocalized bebaviour. When $\delta \gamma$ is large enough, the NHSE dominates and the ground state becomes localized again.

%The NHSE is known as collapsing modes to the boundary, while in this instance, we use it to delocalize the defect mode. Our results indicate that non-Hermiticity offers a new degree of freedom to control the spatial distribution of polaritons.

\begin{figure}
\centering
\includegraphics[width=1\columnwidth]{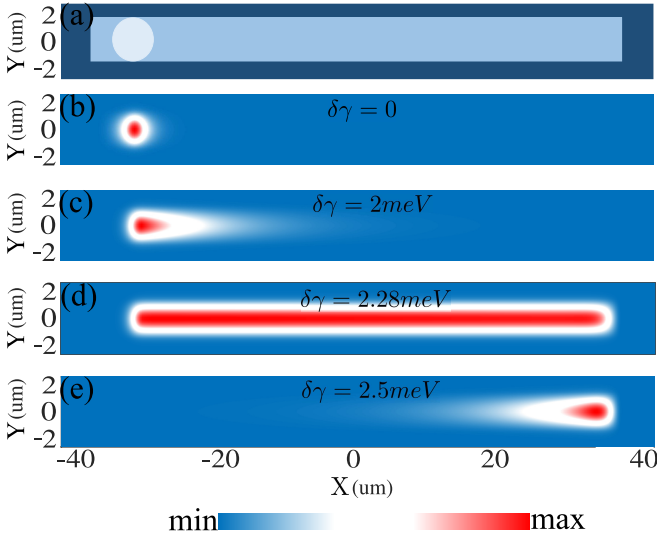}
\caption{(a) The schematic figure of a waveguide with a defect on the left-hand side. (b)$\sim$(e) Spatial distribution of the ground state with different decay rates. Ground state distribution changes drastically from localized inside the defect, dragged into the waveguide, then extended in the waveguide and finally localized at the right end.}
\label{Fig2}
\end{figure}

\emph{Lasing in the extended defect mode.}
The driven-dissipative nature of polaritons offers an excellent platform for studying lasing. With the development of topological polaritons, lasing in the topological edge mode, such as the SSH edge mode in a one-dimensional lattice and the chiral edge state in a two-dimensional lattice, has been proposed and realized~\cite{PhysRevLett.122.083902,St-Jean:2017ws}. In this work, we achieve lasing in the extended defect mode in the presence of the NHSE. By considering the circularly polarized incoherent pump and nonlinear decay, the polariton wavefunction is described by:

\begin{align}
i\hbar \frac{\partial \Psi_{\pm}}{\partial t} & = (1-i\beta) \left[ \left( -\frac{\hbar^2 \nabla^2}{2m}+V \mp 2i\alpha\frac{\partial}{\partial x} \right)\Psi_{\pm} + \Delta_T \Psi_{\mp}  \right] \notag
\\ & + i(P_{\pm}-\gamma)\Psi_{\pm} - i\alpha_1 |\Psi_{\pm}|^2\Psi_{\pm}.
\end{align}

Here, $\beta$ represents energy relaxation ~\cite{Estrecho:2018vv}, $iP_{\pm}$ denotes the circular polarized incoherent pump, and $\alpha_1$ is the nonlinear decay. The relaxation term causes the extended ground state to have the highest gain, which leads to lasing at the desired extended morphed state. To achieve the NHSE, we select the incoherent circularly polarized pump to be spatially uniform with different strengths ($P_+ = 3.3~meV, P_- = 0$), thus creating different effective decay rates. The steady state is shown in Fig. 3(a), where polaritons are distributed almost uniformly throughout the waveguide. Notice that this distribution is similar to the extended state shown in Fig. 2(d). We then calculate the polariton intensity as a function of energy by Fourier transforming the wavefunction $\Psi_{\pm}$ obtained from Eq. (3). Subsequently, we sum up all the intensity along the real space. As shown in Fig. 3(c), the intensity distribution in energy from Fig. 3(a) reveals a peak at around $-3.45~meV$, which is similar to the extended defect state's eigenenergy calculated by diagonalizing the linear Hamiltonian as shown in Fig. 3(b). This further proves that we have obtained polariton lasing in the extended defect state.

\begin{figure}
\centering
\includegraphics[width=1\columnwidth]{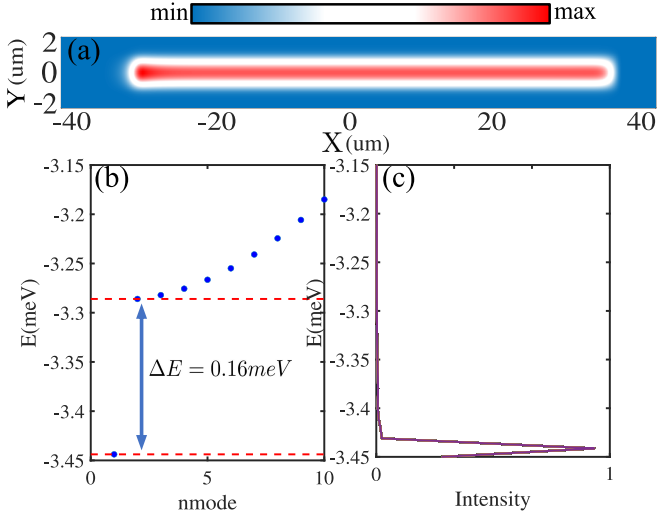}
\caption{(a) The steady state's spatial distribution obtained from Eq. (3). The polaritons are uniformly distributed all over the waveguide like the case of Fig. 2(d). (b) Eigenenergies solved in the linear case ($\alpha_1 = 0$). The energy bandgap between ground state and others is around 0.19 $meV$. (c) Intensity of polaritons as a function of energy. Parameters: $\beta = 0.5, \alpha = 1.7~meV, \Delta_T = 6~meV, \alpha_1 = 1~\mu eV \mu m^2,P_+ = 3.3~meV,P_- = 0$.}
\label{Fig3}
\end{figure}

\emph{Spatial coherence in an extended defect mode.}\label{coherence}
\begin{figure}
\centering
\includegraphics[width=1\columnwidth]{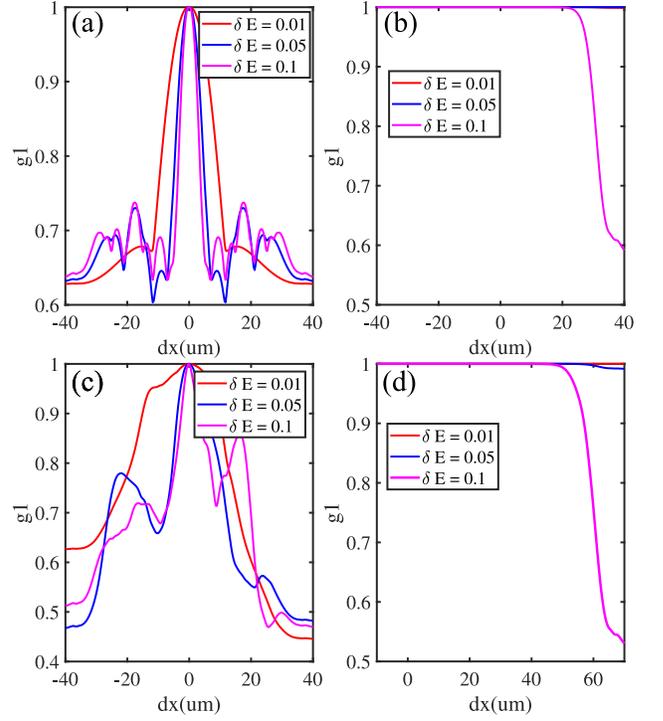}
\caption{(a) and (b) Spatial coherence calculated from Eq. (4-5) in a normal cavity case and the NHSE morphing case under different strengths of fluctuations respectively. (c) and (d) Spatial coherence calculated in presence of same strength of disorder in a normal cavity and the NHSE morphing case.}
\label{Fig4}
\end{figure}
Owing to the deeper potential, the ground state of the defect carries less energy than the ground state of the waveguide. This energy difference creates a bandgap of around 0.2meV. As a result, after the ground state morphing, it is not only distributed all over the waveguide but is also energetically separated from other modes. This unique property enables the possibility of achieving large-area spatial coherence. We calculate the spatial coherence,  $g_1(dx,dy)$ using the formula:
\begin{equation}
g_1(dx,dy) = \frac{\sum_n \Psi_n^*(x,y)\Psi_n(x + dx,y+dy)e^{-\frac{(E_n-E_g)^2}{\delta E^2}}}{\sqrt{I(x,y) I(x+dx,y+dy)}},
\end{equation}
\begin{equation}
I(x,y) = \sum_n |\Psi_n (x,y)|^2 e^{-\frac{(E_n-E_g)^2}{\delta E^2}}.
\end{equation}
This formula is equivalent to that used in~\cite{https://doi.org/10.1002/pssb.201248074}, however, assuming a Gaussian energy distribution function. One might've expected a Bose-Einstein distribution, however, it was shown that this results in a poor fit to experiments given the non-equilibrium nature of polariton systems~\cite{https://doi.org/10.1002/pssb.201248074}.
Here $I(x,y)$ is the intensity profile of polaritons, $E_n$ is the energy of the modes, $E_g$ is the energy of the ground state, $\delta E$ is the energy fluctuation. If $\delta E$ is infinitesimally small corresponding to a perfect condensate, $g_1$ will equal to 1.
% When $\delta E$ increases, we model the excitation of multiple condensate modes within an energy range, and $g_1$ will decrease. 
In the calculation, we take the reference point $(x,y)$ as the point that has the highest intensity.

In a typical cavity, because of the continuous energy spectrum, more than one mode is excited simultaneously, which results in coherence decay, as shown in Fig. 4(a).  It's important to note that with a larger $\delta E$, $g_1$ drops more rapidly. In comparison, the ground state morphing mode displays better coherence all over the waveguide due to the separation of energy and the extended distribution nature. When $\delta E$ is small, almost just the ground state is excited and $g_1$ is nearly one in the whole waveguide. Of course if $\delta E$ is increased too much, coherence will drop, first at the right end of the waveguide. This can be interpreted as other skin modes being excited, which are dominantly localized at the right edge of the waveguide. Note that the coherence remains high throughout the left region of the waveguide.

%One of the topological advantages is the robustness against disorder which naturally present in experiments. So here we consider a random disorder inside the waveguide and compare the spatial coherence in both cases. The strength of disorder considered here is within the bandgap. 
A well-known advantage of topological systems is the robustness of their states with respect to disorder (provided that the disorder strength is within the topological bandgap). In Fig. 4(c) and (d), we compare the spatial coherence in a regular waveguide and our morphing mode system in the presence of disorder (added as a random potential).
%into Eq. (3) .
We notice that the morphing mode still presents high spatial coherence, which shows the robustness to disorder. In comparison, the breakdown of spatial coherence with the same strength of disorder is also shown in Fig. 4(c) for a regular waveguide.

\emph{Conclusion.}
We consider the hybridization of non-Hermitian skin effect modes and a trapped mode in exciton-polariton waveguides. Because of the competition between the trap and NHSE, the localized defect ground state can become extended inside the waveguide. We then show that lasing can be obtained in the extended defect mode. The large bandgap of this ground state and the large area distribution properties make it possible to achieve a significant spatial coherence length, considerably enhanced compared to non-topological systems, especially in the presence of disorder.

\emph{Acknowledgement.}
The work was supported by the Ministry of Education, Singapore (Grant No. MOE-T2EP50121-0020).
\bibliography{main}

\begin{thebibliography}{53}%
\makeatletter
\providecommand \@ifxundefined [1]{%
 \@ifx{#1\undefined}
}%
\providecommand \@ifnum [1]{%
 \ifnum #1\expandafter \@firstoftwo
 \else \expandafter \@secondoftwo
 \fi
}%
\providecommand \@ifx [1]{%
 \ifx #1\expandafter \@firstoftwo
 \else \expandafter \@secondoftwo
 \fi
}%
\providecommand \natexlab [1]{#1}%
\providecommand \enquote  [1]{``#1''}%
\providecommand \bibnamefont  [1]{#1}%
\providecommand \bibfnamefont [1]{#1}%
\providecommand \citenamefont [1]{#1}%
\providecommand \href@noop [0]{\@secondoftwo}%
\providecommand \href [0]{\begingroup \@sanitize@url \@href}%
\providecommand \@href[1]{\@@startlink{#1}\@@href}%
\providecommand \@@href[1]{\endgroup#1\@@endlink}%
\providecommand \@sanitize@url [0]{\catcode `\\12\catcode `\$12\catcode
  `\&12\catcode `\#12\catcode `\^12\catcode `\_12\catcode `\%12\relax}%
\providecommand \@@startlink[1]{}%
\providecommand \@@endlink[0]{}%
\providecommand \url  [0]{\begingroup\@sanitize@url \@url }%
\providecommand \@url [1]{\endgroup\@href {#1}{\urlprefix }}%
\providecommand \urlprefix  [0]{URL }%
\providecommand \Eprint [0]{\href }%
\providecommand \doibase [0]{http://dx.doi.org/}%
\providecommand \selectlanguage [0]{\@gobble}%
\providecommand \bibinfo  [0]{\@secondoftwo}%
\providecommand \bibfield  [0]{\@secondoftwo}%
\providecommand \translation [1]{[#1]}%
\providecommand \BibitemOpen [0]{}%
\providecommand \bibitemStop [0]{}%
\providecommand \bibitemNoStop [0]{.\EOS\space}%
\providecommand \EOS [0]{\spacefactor3000\relax}%
\providecommand \BibitemShut  [1]{\csname bibitem#1\endcsname}%
\let\auto@bib@innerbib\@empty
%</preamble>
\bibitem [{\citenamefont {Deng}\ \emph {et~al.}(2010)\citenamefont {Deng},
  \citenamefont {Haug},\ and\ \citenamefont {Yamamoto}}]{RevModPhys.82.1489}%
  \BibitemOpen
  \bibfield  {author} {\bibinfo {author} {\bibfnamefont {H.}~\bibnamefont
  {Deng}}, \bibinfo {author} {\bibfnamefont {H.}~\bibnamefont {Haug}}, \ and\
  \bibinfo {author} {\bibfnamefont {Y.}~\bibnamefont {Yamamoto}},\ }\href
  {\doibase 10.1103/RevModPhys.82.1489} {\bibfield  {journal} {\bibinfo
  {journal} {Rev. Mod. Phys.}\ }\textbf {\bibinfo {volume} {82}},\ \bibinfo
  {pages} {1489} (\bibinfo {year} {2010})}\BibitemShut {NoStop}%
\bibitem [{\citenamefont {Carusotto}\ and\ \citenamefont
  {Ciuti}(2013)}]{RevModPhys.85.299}%
  \BibitemOpen
  \bibfield  {author} {\bibinfo {author} {\bibfnamefont {I.}~\bibnamefont
  {Carusotto}}\ and\ \bibinfo {author} {\bibfnamefont {C.}~\bibnamefont
  {Ciuti}},\ }\href {\doibase 10.1103/RevModPhys.85.299} {\bibfield  {journal}
  {\bibinfo  {journal} {Rev. Mod. Phys.}\ }\textbf {\bibinfo {volume} {85}},\
  \bibinfo {pages} {299} (\bibinfo {year} {2013})}\BibitemShut {NoStop}%
\bibitem [{\citenamefont {Deng}\ \emph {et~al.}(2002)\citenamefont {Deng},
  \citenamefont {Weihs}, \citenamefont {Santori}, \citenamefont {Bloch},\ and\
  \citenamefont {Yamamoto}}]{doi:10.1126/science.1074464}%
  \BibitemOpen
  \bibfield  {author} {\bibinfo {author} {\bibfnamefont {H.}~\bibnamefont
  {Deng}}, \bibinfo {author} {\bibfnamefont {G.}~\bibnamefont {Weihs}},
  \bibinfo {author} {\bibfnamefont {C.}~\bibnamefont {Santori}}, \bibinfo
  {author} {\bibfnamefont {J.}~\bibnamefont {Bloch}}, \ and\ \bibinfo {author}
  {\bibfnamefont {Y.}~\bibnamefont {Yamamoto}},\ }\href {\doibase
  10.1126/science.1074464} {\bibfield  {journal} {\bibinfo  {journal}
  {Science}\ }\textbf {\bibinfo {volume} {298}},\ \bibinfo {pages} {199}
  (\bibinfo {year} {2002})}\BibitemShut {NoStop}%
\bibitem [{\citenamefont {Christopoulos}\ \emph {et~al.}(2007)\citenamefont
  {Christopoulos}, \citenamefont {von H\"ogersthal}, \citenamefont {Grundy},
  \citenamefont {Lagoudakis}, \citenamefont {Kavokin}, \citenamefont
  {Baumberg}, \citenamefont {Christmann}, \citenamefont {Butt\'e},
  \citenamefont {Feltin}, \citenamefont {Carlin},\ and\ \citenamefont
  {Grandjean}}]{PhysRevLett.98.126405}%
  \BibitemOpen
  \bibfield  {author} {\bibinfo {author} {\bibfnamefont {S.}~\bibnamefont
  {Christopoulos}}, \bibinfo {author} {\bibfnamefont {G.~B.~H.}\ \bibnamefont
  {von H\"ogersthal}}, \bibinfo {author} {\bibfnamefont {A.~J.~D.}\
  \bibnamefont {Grundy}}, \bibinfo {author} {\bibfnamefont {P.~G.}\
  \bibnamefont {Lagoudakis}}, \bibinfo {author} {\bibfnamefont {A.~V.}\
  \bibnamefont {Kavokin}}, \bibinfo {author} {\bibfnamefont {J.~J.}\
  \bibnamefont {Baumberg}}, \bibinfo {author} {\bibfnamefont {G.}~\bibnamefont
  {Christmann}}, \bibinfo {author} {\bibfnamefont {R.}~\bibnamefont {Butt\'e}},
  \bibinfo {author} {\bibfnamefont {E.}~\bibnamefont {Feltin}}, \bibinfo
  {author} {\bibfnamefont {J.-F.}\ \bibnamefont {Carlin}}, \ and\ \bibinfo
  {author} {\bibfnamefont {N.}~\bibnamefont {Grandjean}},\ }\href {\doibase
  10.1103/PhysRevLett.98.126405} {\bibfield  {journal} {\bibinfo  {journal}
  {Phys. Rev. Lett.}\ }\textbf {\bibinfo {volume} {98}},\ \bibinfo {pages}
  {126405} (\bibinfo {year} {2007})}\BibitemShut {NoStop}%
\bibitem [{\citenamefont {K{\'e}na-Cohen}\ and\ \citenamefont
  {Forrest}(2010)}]{Kena-Cohen:2010ue}%
  \BibitemOpen
  \bibfield  {author} {\bibinfo {author} {\bibfnamefont {S.}~\bibnamefont
  {K{\'e}na-Cohen}}\ and\ \bibinfo {author} {\bibfnamefont {S.~R.}\
  \bibnamefont {Forrest}},\ }\href {\doibase 10.1038/nphoton.2010.86}
  {\bibfield  {journal} {\bibinfo  {journal} {Nature Photonics}\ }\textbf
  {\bibinfo {volume} {4}},\ \bibinfo {pages} {371} (\bibinfo {year}
  {2010})}\BibitemShut {NoStop}%
\bibitem [{\citenamefont {Su}\ \emph {et~al.}(2017)\citenamefont {Su},
  \citenamefont {Diederichs}, \citenamefont {Wang}, \citenamefont {Liew},
  \citenamefont {Zhao}, \citenamefont {Liu}, \citenamefont {Xu}, \citenamefont
  {Chen},\ and\ \citenamefont {Xiong}}]{doi:10.1021/acs.nanolett.7b01956}%
  \BibitemOpen
  \bibfield  {author} {\bibinfo {author} {\bibfnamefont {R.}~\bibnamefont
  {Su}}, \bibinfo {author} {\bibfnamefont {C.}~\bibnamefont {Diederichs}},
  \bibinfo {author} {\bibfnamefont {J.}~\bibnamefont {Wang}}, \bibinfo {author}
  {\bibfnamefont {T.~C.~H.}\ \bibnamefont {Liew}}, \bibinfo {author}
  {\bibfnamefont {J.}~\bibnamefont {Zhao}}, \bibinfo {author} {\bibfnamefont
  {S.}~\bibnamefont {Liu}}, \bibinfo {author} {\bibfnamefont {W.}~\bibnamefont
  {Xu}}, \bibinfo {author} {\bibfnamefont {Z.}~\bibnamefont {Chen}}, \ and\
  \bibinfo {author} {\bibfnamefont {Q.}~\bibnamefont {Xiong}},\ }\href
  {\doibase 10.1021/acs.nanolett.7b01956} {\bibfield  {journal} {\bibinfo
  {journal} {Nano Letters}\ }\textbf {\bibinfo {volume} {17}},\ \bibinfo
  {pages} {3982} (\bibinfo {year} {2017})}\BibitemShut {NoStop}%
\bibitem [{\citenamefont {Su}\ \emph {et~al.}(2020)\citenamefont {Su},
  \citenamefont {Ghosh}, \citenamefont {Wang}, \citenamefont {Liu},
  \citenamefont {Diederichs}, \citenamefont {Liew},\ and\ \citenamefont
  {Xiong}}]{Su:2020wu}%
  \BibitemOpen
  \bibfield  {author} {\bibinfo {author} {\bibfnamefont {R.}~\bibnamefont
  {Su}}, \bibinfo {author} {\bibfnamefont {S.}~\bibnamefont {Ghosh}}, \bibinfo
  {author} {\bibfnamefont {J.}~\bibnamefont {Wang}}, \bibinfo {author}
  {\bibfnamefont {S.}~\bibnamefont {Liu}}, \bibinfo {author} {\bibfnamefont
  {C.}~\bibnamefont {Diederichs}}, \bibinfo {author} {\bibfnamefont {T.~C.~H.}\
  \bibnamefont {Liew}}, \ and\ \bibinfo {author} {\bibfnamefont
  {Q.}~\bibnamefont {Xiong}},\ }\href {\doibase 10.1038/s41567-019-0764-5}
  {\bibfield  {journal} {\bibinfo  {journal} {Nature Physics}\ }\textbf
  {\bibinfo {volume} {16}},\ \bibinfo {pages} {301} (\bibinfo {year}
  {2020})}\BibitemShut {NoStop}%
\bibitem [{\citenamefont {Harder}\ \emph {et~al.}(2021)\citenamefont {Harder},
  \citenamefont {Sun}, \citenamefont {Egorov}, \citenamefont {Vakulchyk},
  \citenamefont {Beierlein}, \citenamefont {Gagel}, \citenamefont {Emmerling},
  \citenamefont {Schneider}, \citenamefont {Peschel}, \citenamefont {Savenko},
  \citenamefont {Klembt},\ and\ \citenamefont
  {H{\"o}fling}}]{doi:10.1021/acsphotonics.0c01958}%
  \BibitemOpen
  \bibfield  {author} {\bibinfo {author} {\bibfnamefont {T.~H.}\ \bibnamefont
  {Harder}}, \bibinfo {author} {\bibfnamefont {M.}~\bibnamefont {Sun}},
  \bibinfo {author} {\bibfnamefont {O.~A.}\ \bibnamefont {Egorov}}, \bibinfo
  {author} {\bibfnamefont {I.}~\bibnamefont {Vakulchyk}}, \bibinfo {author}
  {\bibfnamefont {J.}~\bibnamefont {Beierlein}}, \bibinfo {author}
  {\bibfnamefont {P.}~\bibnamefont {Gagel}}, \bibinfo {author} {\bibfnamefont
  {M.}~\bibnamefont {Emmerling}}, \bibinfo {author} {\bibfnamefont
  {C.}~\bibnamefont {Schneider}}, \bibinfo {author} {\bibfnamefont
  {U.}~\bibnamefont {Peschel}}, \bibinfo {author} {\bibfnamefont {I.~G.}\
  \bibnamefont {Savenko}}, \bibinfo {author} {\bibfnamefont {S.}~\bibnamefont
  {Klembt}}, \ and\ \bibinfo {author} {\bibfnamefont {S.}~\bibnamefont
  {H{\"o}fling}},\ }\href {\doibase 10.1021/acsphotonics.0c01958} {\bibfield
  {journal} {\bibinfo  {journal} {ACS Photonics}\ }\textbf {\bibinfo {volume}
  {8}},\ \bibinfo {pages} {1377} (\bibinfo {year} {2021})}\BibitemShut
  {NoStop}%
\bibitem [{\citenamefont {Su}\ \emph {et~al.}(2021{\natexlab{a}})\citenamefont
  {Su}, \citenamefont {Ghosh}, \citenamefont {Liew},\ and\ \citenamefont
  {Xiong}}]{doi:10.1126/sciadv.abf8049}%
  \BibitemOpen
  \bibfield  {author} {\bibinfo {author} {\bibfnamefont {R.}~\bibnamefont
  {Su}}, \bibinfo {author} {\bibfnamefont {S.}~\bibnamefont {Ghosh}}, \bibinfo
  {author} {\bibfnamefont {T.~C.~H.}\ \bibnamefont {Liew}}, \ and\ \bibinfo
  {author} {\bibfnamefont {Q.}~\bibnamefont {Xiong}},\ }\href {\doibase
  10.1126/sciadv.abf8049} {\bibfield  {journal} {\bibinfo  {journal} {Science
  Advances}\ }\textbf {\bibinfo {volume} {7}},\ \bibinfo {pages} {eabf8049}
  (\bibinfo {year} {2021}{\natexlab{a}})}\BibitemShut {NoStop}%
\bibitem [{\citenamefont {Kokhanchik}\ \emph {et~al.}(2022)\citenamefont
  {Kokhanchik}, \citenamefont {Solnyshkov}, \citenamefont {St\"oferle},
  \citenamefont {Pi\ifmmode~\mbox{\k{e}}\else \k{e}\fi{}tka}, \citenamefont
  {Szczytko},\ and\ \citenamefont {Malpuech}}]{PhysRevLett.129.246801}%
  \BibitemOpen
  \bibfield  {author} {\bibinfo {author} {\bibfnamefont {P.}~\bibnamefont
  {Kokhanchik}}, \bibinfo {author} {\bibfnamefont {D.}~\bibnamefont
  {Solnyshkov}}, \bibinfo {author} {\bibfnamefont {T.}~\bibnamefont
  {St\"oferle}}, \bibinfo {author} {\bibfnamefont {B.}~\bibnamefont
  {Pi\ifmmode~\mbox{\k{e}}\else \k{e}\fi{}tka}}, \bibinfo {author}
  {\bibfnamefont {J.}~\bibnamefont {Szczytko}}, \ and\ \bibinfo {author}
  {\bibfnamefont {G.}~\bibnamefont {Malpuech}},\ }\href {\doibase
  10.1103/PhysRevLett.129.246801} {\bibfield  {journal} {\bibinfo  {journal}
  {Phys. Rev. Lett.}\ }\textbf {\bibinfo {volume} {129}},\ \bibinfo {pages}
  {246801} (\bibinfo {year} {2022})}\BibitemShut {NoStop}%
\bibitem [{\citenamefont {Bardyn}\ \emph {et~al.}(2015)\citenamefont {Bardyn},
  \citenamefont {Karzig}, \citenamefont {Refael},\ and\ \citenamefont
  {Liew}}]{PhysRevB.91.161413}%
  \BibitemOpen
  \bibfield  {author} {\bibinfo {author} {\bibfnamefont {C.-E.}\ \bibnamefont
  {Bardyn}}, \bibinfo {author} {\bibfnamefont {T.}~\bibnamefont {Karzig}},
  \bibinfo {author} {\bibfnamefont {G.}~\bibnamefont {Refael}}, \ and\ \bibinfo
  {author} {\bibfnamefont {T.~C.~H.}\ \bibnamefont {Liew}},\ }\href {\doibase
  10.1103/PhysRevB.91.161413} {\bibfield  {journal} {\bibinfo  {journal} {Phys.
  Rev. B}\ }\textbf {\bibinfo {volume} {91}},\ \bibinfo {pages} {161413}
  (\bibinfo {year} {2015})}\BibitemShut {NoStop}%
\bibitem [{\citenamefont {Nalitov}\ \emph {et~al.}(2015)\citenamefont
  {Nalitov}, \citenamefont {Solnyshkov},\ and\ \citenamefont
  {Malpuech}}]{PhysRevLett.114.116401}%
  \BibitemOpen
  \bibfield  {author} {\bibinfo {author} {\bibfnamefont {A.~V.}\ \bibnamefont
  {Nalitov}}, \bibinfo {author} {\bibfnamefont {D.~D.}\ \bibnamefont
  {Solnyshkov}}, \ and\ \bibinfo {author} {\bibfnamefont {G.}~\bibnamefont
  {Malpuech}},\ }\href {\doibase 10.1103/PhysRevLett.114.116401} {\bibfield
  {journal} {\bibinfo  {journal} {Phys. Rev. Lett.}\ }\textbf {\bibinfo
  {volume} {114}},\ \bibinfo {pages} {116401} (\bibinfo {year}
  {2015})}\BibitemShut {NoStop}%
\bibitem [{\citenamefont {Klembt}\ \emph {et~al.}(2018)\citenamefont {Klembt},
  \citenamefont {Harder}, \citenamefont {Egorov}, \citenamefont {Winkler},
  \citenamefont {Ge}, \citenamefont {Bandres}, \citenamefont {Emmerling},
  \citenamefont {Worschech}, \citenamefont {Liew}, \citenamefont {Segev},
  \citenamefont {Schneider},\ and\ \citenamefont
  {H{\"o}fling}}]{Klembt:2018wu}%
  \BibitemOpen
  \bibfield  {author} {\bibinfo {author} {\bibfnamefont {S.}~\bibnamefont
  {Klembt}}, \bibinfo {author} {\bibfnamefont {T.~H.}\ \bibnamefont {Harder}},
  \bibinfo {author} {\bibfnamefont {O.~A.}\ \bibnamefont {Egorov}}, \bibinfo
  {author} {\bibfnamefont {K.}~\bibnamefont {Winkler}}, \bibinfo {author}
  {\bibfnamefont {R.}~\bibnamefont {Ge}}, \bibinfo {author} {\bibfnamefont
  {M.~A.}\ \bibnamefont {Bandres}}, \bibinfo {author} {\bibfnamefont
  {M.}~\bibnamefont {Emmerling}}, \bibinfo {author} {\bibfnamefont
  {L.}~\bibnamefont {Worschech}}, \bibinfo {author} {\bibfnamefont {T.~C.~H.}\
  \bibnamefont {Liew}}, \bibinfo {author} {\bibfnamefont {M.}~\bibnamefont
  {Segev}}, \bibinfo {author} {\bibfnamefont {C.}~\bibnamefont {Schneider}}, \
  and\ \bibinfo {author} {\bibfnamefont {S.}~\bibnamefont {H{\"o}fling}},\
  }\href {\doibase 10.1038/s41586-018-0601-5} {\bibfield  {journal} {\bibinfo
  {journal} {Nature}\ }\textbf {\bibinfo {volume} {562}},\ \bibinfo {pages}
  {552} (\bibinfo {year} {2018})}\BibitemShut {NoStop}%
\bibitem [{\citenamefont {Bleu}\ \emph {et~al.}(2017)\citenamefont {Bleu},
  \citenamefont {Solnyshkov},\ and\ \citenamefont
  {Malpuech}}]{PhysRevB.96.165432}%
  \BibitemOpen
  \bibfield  {author} {\bibinfo {author} {\bibfnamefont {O.}~\bibnamefont
  {Bleu}}, \bibinfo {author} {\bibfnamefont {D.~D.}\ \bibnamefont
  {Solnyshkov}}, \ and\ \bibinfo {author} {\bibfnamefont {G.}~\bibnamefont
  {Malpuech}},\ }\href {\doibase 10.1103/PhysRevB.96.165432} {\bibfield
  {journal} {\bibinfo  {journal} {Phys. Rev. B}\ }\textbf {\bibinfo {volume}
  {96}},\ \bibinfo {pages} {165432} (\bibinfo {year} {2017})}\BibitemShut
  {NoStop}%
\bibitem [{\citenamefont {Liu}\ \emph {et~al.}(2020)\citenamefont {Liu},
  \citenamefont {Ji}, \citenamefont {Wang}, \citenamefont {Modi}, \citenamefont
  {Hwang}, \citenamefont {Zheng}, \citenamefont {Sorger}, \citenamefont {Pan},\
  and\ \citenamefont {Agarwal}}]{doi:10.1126/science.abc4975}%
  \BibitemOpen
  \bibfield  {author} {\bibinfo {author} {\bibfnamefont {W.}~\bibnamefont
  {Liu}}, \bibinfo {author} {\bibfnamefont {Z.}~\bibnamefont {Ji}}, \bibinfo
  {author} {\bibfnamefont {Y.}~\bibnamefont {Wang}}, \bibinfo {author}
  {\bibfnamefont {G.}~\bibnamefont {Modi}}, \bibinfo {author} {\bibfnamefont
  {M.}~\bibnamefont {Hwang}}, \bibinfo {author} {\bibfnamefont
  {B.}~\bibnamefont {Zheng}}, \bibinfo {author} {\bibfnamefont {V.~J.}\
  \bibnamefont {Sorger}}, \bibinfo {author} {\bibfnamefont {A.}~\bibnamefont
  {Pan}}, \ and\ \bibinfo {author} {\bibfnamefont {R.}~\bibnamefont
  {Agarwal}},\ }\href {\doibase 10.1126/science.abc4975} {\bibfield  {journal}
  {\bibinfo  {journal} {Science}\ }\textbf {\bibinfo {volume} {370}},\ \bibinfo
  {pages} {600} (\bibinfo {year} {2020})}\BibitemShut {NoStop}%
\bibitem [{\citenamefont {Banerjee}\ \emph {et~al.}(2021)\citenamefont
  {Banerjee}, \citenamefont {Mandal},\ and\ \citenamefont
  {Liew}}]{PhysRevB.103.L201406}%
  \BibitemOpen
  \bibfield  {author} {\bibinfo {author} {\bibfnamefont {R.}~\bibnamefont
  {Banerjee}}, \bibinfo {author} {\bibfnamefont {S.}~\bibnamefont {Mandal}}, \
  and\ \bibinfo {author} {\bibfnamefont {T.~C.~H.}\ \bibnamefont {Liew}},\
  }\href {\doibase 10.1103/PhysRevB.103.L201406} {\bibfield  {journal}
  {\bibinfo  {journal} {Phys. Rev. B}\ }\textbf {\bibinfo {volume} {103}},\
  \bibinfo {pages} {L201406} (\bibinfo {year} {2021})}\BibitemShut {NoStop}%
\bibitem [{\citenamefont {Mandal}\ \emph {et~al.}(2019)\citenamefont {Mandal},
  \citenamefont {Ge},\ and\ \citenamefont {Liew}}]{PhysRevB.99.115423}%
  \BibitemOpen
  \bibfield  {author} {\bibinfo {author} {\bibfnamefont {S.}~\bibnamefont
  {Mandal}}, \bibinfo {author} {\bibfnamefont {R.}~\bibnamefont {Ge}}, \ and\
  \bibinfo {author} {\bibfnamefont {T.~C.~H.}\ \bibnamefont {Liew}},\ }\href
  {\doibase 10.1103/PhysRevB.99.115423} {\bibfield  {journal} {\bibinfo
  {journal} {Phys. Rev. B}\ }\textbf {\bibinfo {volume} {99}},\ \bibinfo
  {pages} {115423} (\bibinfo {year} {2019})}\BibitemShut {NoStop}%
\bibitem [{\citenamefont {Bao}\ \emph {et~al.}(2022)\citenamefont {Bao},
  \citenamefont {Mandal}, \citenamefont {Xu}, \citenamefont {Xu}, \citenamefont
  {Banerjee},\ and\ \citenamefont {Liew}}]{PhysRevB.106.235310}%
  \BibitemOpen
  \bibfield  {author} {\bibinfo {author} {\bibfnamefont {R.}~\bibnamefont
  {Bao}}, \bibinfo {author} {\bibfnamefont {S.}~\bibnamefont {Mandal}},
  \bibinfo {author} {\bibfnamefont {H.}~\bibnamefont {Xu}}, \bibinfo {author}
  {\bibfnamefont {X.}~\bibnamefont {Xu}}, \bibinfo {author} {\bibfnamefont
  {R.}~\bibnamefont {Banerjee}}, \ and\ \bibinfo {author} {\bibfnamefont
  {T.~C.~H.}\ \bibnamefont {Liew}},\ }\href {\doibase
  10.1103/PhysRevB.106.235310} {\bibfield  {journal} {\bibinfo  {journal}
  {Phys. Rev. B}\ }\textbf {\bibinfo {volume} {106}},\ \bibinfo {pages}
  {235310} (\bibinfo {year} {2022})}\BibitemShut {NoStop}%
\bibitem [{\citenamefont {Banerjee}\ \emph {et~al.}(2020)\citenamefont
  {Banerjee}, \citenamefont {Mandal},\ and\ \citenamefont
  {Liew}}]{PhysRevLett.124.063901}%
  \BibitemOpen
  \bibfield  {author} {\bibinfo {author} {\bibfnamefont {R.}~\bibnamefont
  {Banerjee}}, \bibinfo {author} {\bibfnamefont {S.}~\bibnamefont {Mandal}}, \
  and\ \bibinfo {author} {\bibfnamefont {T.~C.~H.}\ \bibnamefont {Liew}},\
  }\href {\doibase 10.1103/PhysRevLett.124.063901} {\bibfield  {journal}
  {\bibinfo  {journal} {Phys. Rev. Lett.}\ }\textbf {\bibinfo {volume} {124}},\
  \bibinfo {pages} {063901} (\bibinfo {year} {2020})}\BibitemShut {NoStop}%
\bibitem [{\citenamefont {Wu}\ \emph {et~al.}(2023)\citenamefont {Wu},
  \citenamefont {Ghosh}, \citenamefont {Gan}, \citenamefont {Shi},
  \citenamefont {Mandal}, \citenamefont {Sun}, \citenamefont {Zhang},
  \citenamefont {Liew}, \citenamefont {Su},\ and\ \citenamefont
  {Xiong}}]{doi:10.1126/sciadv.adg4322}%
  \BibitemOpen
  \bibfield  {author} {\bibinfo {author} {\bibfnamefont {J.}~\bibnamefont
  {Wu}}, \bibinfo {author} {\bibfnamefont {S.}~\bibnamefont {Ghosh}}, \bibinfo
  {author} {\bibfnamefont {Y.}~\bibnamefont {Gan}}, \bibinfo {author}
  {\bibfnamefont {Y.}~\bibnamefont {Shi}}, \bibinfo {author} {\bibfnamefont
  {S.}~\bibnamefont {Mandal}}, \bibinfo {author} {\bibfnamefont
  {H.}~\bibnamefont {Sun}}, \bibinfo {author} {\bibfnamefont {B.}~\bibnamefont
  {Zhang}}, \bibinfo {author} {\bibfnamefont {T.~C.~H.}\ \bibnamefont {Liew}},
  \bibinfo {author} {\bibfnamefont {R.}~\bibnamefont {Su}}, \ and\ \bibinfo
  {author} {\bibfnamefont {Q.}~\bibnamefont {Xiong}},\ }\href {\doibase
  10.1126/sciadv.adg4322} {\bibfield  {journal} {\bibinfo  {journal} {Science
  Advances}\ }\textbf {\bibinfo {volume} {9}},\ \bibinfo {pages} {eadg4322}
  (\bibinfo {year} {2023})}\BibitemShut {NoStop}%
\bibitem [{\citenamefont {Zhang}\ \emph {et~al.}(2020)\citenamefont {Zhang},
  \citenamefont {Kartashov}, \citenamefont {Torner}, \citenamefont {Li},\ and\
  \citenamefont {Ferrando}}]{Zhang:20}%
  \BibitemOpen
  \bibfield  {author} {\bibinfo {author} {\bibfnamefont {Y.}~\bibnamefont
  {Zhang}}, \bibinfo {author} {\bibfnamefont {Y.~V.}\ \bibnamefont
  {Kartashov}}, \bibinfo {author} {\bibfnamefont {L.}~\bibnamefont {Torner}},
  \bibinfo {author} {\bibfnamefont {Y.}~\bibnamefont {Li}}, \ and\ \bibinfo
  {author} {\bibfnamefont {A.}~\bibnamefont {Ferrando}},\ }\href {\doibase
  10.1364/OL.396039} {\bibfield  {journal} {\bibinfo  {journal} {Opt. Lett.}\
  }\textbf {\bibinfo {volume} {45}},\ \bibinfo {pages} {4710} (\bibinfo {year}
  {2020})}\BibitemShut {NoStop}%
\bibitem [{\citenamefont {Gao}\ \emph {et~al.}(2015)\citenamefont {Gao},
  \citenamefont {Estrecho}, \citenamefont {Bliokh}, \citenamefont {Liew},
  \citenamefont {Fraser}, \citenamefont {Brodbeck}, \citenamefont {Kamp},
  \citenamefont {Schneider}, \citenamefont {H{\"o}fling}, \citenamefont
  {Yamamoto}, \citenamefont {Nori}, \citenamefont {Kivshar}, \citenamefont
  {Truscott}, \citenamefont {Dall},\ and\ \citenamefont
  {Ostrovskaya}}]{Gao:2015uy}%
  \BibitemOpen
  \bibfield  {author} {\bibinfo {author} {\bibfnamefont {T.}~\bibnamefont
  {Gao}}, \bibinfo {author} {\bibfnamefont {E.}~\bibnamefont {Estrecho}},
  \bibinfo {author} {\bibfnamefont {K.~Y.}\ \bibnamefont {Bliokh}}, \bibinfo
  {author} {\bibfnamefont {T.~C.~H.}\ \bibnamefont {Liew}}, \bibinfo {author}
  {\bibfnamefont {M.~D.}\ \bibnamefont {Fraser}}, \bibinfo {author}
  {\bibfnamefont {S.}~\bibnamefont {Brodbeck}}, \bibinfo {author}
  {\bibfnamefont {M.}~\bibnamefont {Kamp}}, \bibinfo {author} {\bibfnamefont
  {C.}~\bibnamefont {Schneider}}, \bibinfo {author} {\bibfnamefont
  {S.}~\bibnamefont {H{\"o}fling}}, \bibinfo {author} {\bibfnamefont
  {Y.}~\bibnamefont {Yamamoto}}, \bibinfo {author} {\bibfnamefont
  {F.}~\bibnamefont {Nori}}, \bibinfo {author} {\bibfnamefont {Y.~S.}\
  \bibnamefont {Kivshar}}, \bibinfo {author} {\bibfnamefont {A.~G.}\
  \bibnamefont {Truscott}}, \bibinfo {author} {\bibfnamefont {R.~G.}\
  \bibnamefont {Dall}}, \ and\ \bibinfo {author} {\bibfnamefont {E.~A.}\
  \bibnamefont {Ostrovskaya}},\ }\href {\doibase 10.1038/nature15522}
  {\bibfield  {journal} {\bibinfo  {journal} {Nature}\ }\textbf {\bibinfo
  {volume} {526}},\ \bibinfo {pages} {554} (\bibinfo {year}
  {2015})}\BibitemShut {NoStop}%
\bibitem [{\citenamefont {Gao}\ \emph {et~al.}(2018)\citenamefont {Gao},
  \citenamefont {Li}, \citenamefont {Bamba},\ and\ \citenamefont
  {Kono}}]{Gao:2018vv}%
  \BibitemOpen
  \bibfield  {author} {\bibinfo {author} {\bibfnamefont {W.}~\bibnamefont
  {Gao}}, \bibinfo {author} {\bibfnamefont {X.}~\bibnamefont {Li}}, \bibinfo
  {author} {\bibfnamefont {M.}~\bibnamefont {Bamba}}, \ and\ \bibinfo {author}
  {\bibfnamefont {J.}~\bibnamefont {Kono}},\ }\href {\doibase
  10.1038/s41566-018-0157-9} {\bibfield  {journal} {\bibinfo  {journal} {Nature
  Photonics}\ }\textbf {\bibinfo {volume} {12}},\ \bibinfo {pages} {362}
  (\bibinfo {year} {2018})}\BibitemShut {NoStop}%
\bibitem [{\citenamefont {Su}\ \emph {et~al.}(2021{\natexlab{b}})\citenamefont
  {Su}, \citenamefont {Estrecho}, \citenamefont {Biega{\'n}ska}, \citenamefont
  {Huang}, \citenamefont {Wurdack}, \citenamefont {Pieczarka}, \citenamefont
  {Truscott}, \citenamefont {Liew}, \citenamefont {Ostrovskaya},\ and\
  \citenamefont {Xiong}}]{doi:10.1126/sciadv.abj8905}%
  \BibitemOpen
  \bibfield  {author} {\bibinfo {author} {\bibfnamefont {R.}~\bibnamefont
  {Su}}, \bibinfo {author} {\bibfnamefont {E.}~\bibnamefont {Estrecho}},
  \bibinfo {author} {\bibfnamefont {D.}~\bibnamefont {Biega{\'n}ska}}, \bibinfo
  {author} {\bibfnamefont {Y.}~\bibnamefont {Huang}}, \bibinfo {author}
  {\bibfnamefont {M.}~\bibnamefont {Wurdack}}, \bibinfo {author} {\bibfnamefont
  {M.}~\bibnamefont {Pieczarka}}, \bibinfo {author} {\bibfnamefont {A.~G.}\
  \bibnamefont {Truscott}}, \bibinfo {author} {\bibfnamefont {T.~C.~H.}\
  \bibnamefont {Liew}}, \bibinfo {author} {\bibfnamefont {E.~A.}\ \bibnamefont
  {Ostrovskaya}}, \ and\ \bibinfo {author} {\bibfnamefont {Q.}~\bibnamefont
  {Xiong}},\ }\href {\doibase 10.1126/sciadv.abj8905} {\bibfield  {journal}
  {\bibinfo  {journal} {Science Advances}\ }\textbf {\bibinfo {volume} {7}},\
  \bibinfo {pages} {eabj8905} (\bibinfo {year}
  {2021}{\natexlab{b}})}\BibitemShut {NoStop}%
\bibitem [{\citenamefont {Dang}\ \emph {et~al.}(2022)\citenamefont {Dang},
  \citenamefont {Zanotti}, \citenamefont {Drouard}, \citenamefont {Chevalier},
  \citenamefont {Tripp{\'e}-Allard}, \citenamefont {Amara}, \citenamefont
  {Deleporte}, \citenamefont {Ardizzone}, \citenamefont {Sanvitto},
  \citenamefont {Andreani}, \citenamefont {Seassal}, \citenamefont {Gerace},\
  and\ \citenamefont {Nguyen}}]{https://doi.org/10.1002/adom.202102386}%
  \BibitemOpen
  \bibfield  {author} {\bibinfo {author} {\bibfnamefont {N.~H.~M.}\
  \bibnamefont {Dang}}, \bibinfo {author} {\bibfnamefont {S.}~\bibnamefont
  {Zanotti}}, \bibinfo {author} {\bibfnamefont {E.}~\bibnamefont {Drouard}},
  \bibinfo {author} {\bibfnamefont {C.}~\bibnamefont {Chevalier}}, \bibinfo
  {author} {\bibfnamefont {G.}~\bibnamefont {Tripp{\'e}-Allard}}, \bibinfo
  {author} {\bibfnamefont {M.}~\bibnamefont {Amara}}, \bibinfo {author}
  {\bibfnamefont {E.}~\bibnamefont {Deleporte}}, \bibinfo {author}
  {\bibfnamefont {V.}~\bibnamefont {Ardizzone}}, \bibinfo {author}
  {\bibfnamefont {D.}~\bibnamefont {Sanvitto}}, \bibinfo {author}
  {\bibfnamefont {L.~C.}\ \bibnamefont {Andreani}}, \bibinfo {author}
  {\bibfnamefont {C.}~\bibnamefont {Seassal}}, \bibinfo {author} {\bibfnamefont
  {D.}~\bibnamefont {Gerace}}, \ and\ \bibinfo {author} {\bibfnamefont {H.~S.}\
  \bibnamefont {Nguyen}},\ }\href {\doibase
  https://doi.org/10.1002/adom.202102386} {\bibfield  {journal} {\bibinfo
  {journal} {Advanced Optical Materials}\ }\textbf {\bibinfo {volume} {10}},\
  \bibinfo {pages} {2102386} (\bibinfo {year} {2022})}\BibitemShut {NoStop}%
\bibitem [{\citenamefont {Mandal}\ \emph {et~al.}(2020)\citenamefont {Mandal},
  \citenamefont {Banerjee}, \citenamefont {Ostrovskaya},\ and\ \citenamefont
  {Liew}}]{PhysRevLett.125.123902}%
  \BibitemOpen
  \bibfield  {author} {\bibinfo {author} {\bibfnamefont {S.}~\bibnamefont
  {Mandal}}, \bibinfo {author} {\bibfnamefont {R.}~\bibnamefont {Banerjee}},
  \bibinfo {author} {\bibfnamefont {E.~A.}\ \bibnamefont {Ostrovskaya}}, \ and\
  \bibinfo {author} {\bibfnamefont {T.~C.~H.}\ \bibnamefont {Liew}},\ }\href
  {\doibase 10.1103/PhysRevLett.125.123902} {\bibfield  {journal} {\bibinfo
  {journal} {Phys. Rev. Lett.}\ }\textbf {\bibinfo {volume} {125}},\ \bibinfo
  {pages} {123902} (\bibinfo {year} {2020})}\BibitemShut {NoStop}%
\bibitem [{\citenamefont {Mandal}\ \emph {et~al.}(2022)\citenamefont {Mandal},
  \citenamefont {Banerjee},\ and\ \citenamefont
  {Liew}}]{doi:10.1021/acsphotonics.1c01425}%
  \BibitemOpen
  \bibfield  {author} {\bibinfo {author} {\bibfnamefont {S.}~\bibnamefont
  {Mandal}}, \bibinfo {author} {\bibfnamefont {R.}~\bibnamefont {Banerjee}}, \
  and\ \bibinfo {author} {\bibfnamefont {T.~C.~H.}\ \bibnamefont {Liew}},\
  }\href {\doibase 10.1021/acsphotonics.1c01425} {\bibfield  {journal}
  {\bibinfo  {journal} {ACS Photonics}\ }\textbf {\bibinfo {volume} {9}},\
  \bibinfo {pages} {527} (\bibinfo {year} {2022})}\BibitemShut {NoStop}%
\bibitem [{\citenamefont {Xu}\ \emph {et~al.}(2021{\natexlab{a}})\citenamefont
  {Xu}, \citenamefont {Dini}, \citenamefont {Xu}, \citenamefont {Banerjee},
  \citenamefont {Mandal},\ and\ \citenamefont {Liew}}]{PhysRevB.104.195301}%
  \BibitemOpen
  \bibfield  {author} {\bibinfo {author} {\bibfnamefont {H.}~\bibnamefont
  {Xu}}, \bibinfo {author} {\bibfnamefont {K.}~\bibnamefont {Dini}}, \bibinfo
  {author} {\bibfnamefont {X.}~\bibnamefont {Xu}}, \bibinfo {author}
  {\bibfnamefont {R.}~\bibnamefont {Banerjee}}, \bibinfo {author}
  {\bibfnamefont {S.}~\bibnamefont {Mandal}}, \ and\ \bibinfo {author}
  {\bibfnamefont {T.~C.~H.}\ \bibnamefont {Liew}},\ }\href {\doibase
  10.1103/PhysRevB.104.195301} {\bibfield  {journal} {\bibinfo  {journal}
  {Phys. Rev. B}\ }\textbf {\bibinfo {volume} {104}},\ \bibinfo {pages}
  {195301} (\bibinfo {year} {2021}{\natexlab{a}})}\BibitemShut {NoStop}%
\bibitem [{\citenamefont {Xu}\ \emph {et~al.}(2021{\natexlab{b}})\citenamefont
  {Xu}, \citenamefont {Xu}, \citenamefont {Mandal}, \citenamefont {Banerjee},
  \citenamefont {Ghosh},\ and\ \citenamefont {Liew}}]{PhysRevB.103.235306}%
  \BibitemOpen
  \bibfield  {author} {\bibinfo {author} {\bibfnamefont {X.}~\bibnamefont
  {Xu}}, \bibinfo {author} {\bibfnamefont {H.}~\bibnamefont {Xu}}, \bibinfo
  {author} {\bibfnamefont {S.}~\bibnamefont {Mandal}}, \bibinfo {author}
  {\bibfnamefont {R.}~\bibnamefont {Banerjee}}, \bibinfo {author}
  {\bibfnamefont {S.}~\bibnamefont {Ghosh}}, \ and\ \bibinfo {author}
  {\bibfnamefont {T.~C.~H.}\ \bibnamefont {Liew}},\ }\href {\doibase
  10.1103/PhysRevB.103.235306} {\bibfield  {journal} {\bibinfo  {journal}
  {Phys. Rev. B}\ }\textbf {\bibinfo {volume} {103}},\ \bibinfo {pages}
  {235306} (\bibinfo {year} {2021}{\natexlab{b}})}\BibitemShut {NoStop}%
\bibitem [{\citenamefont {Xu}\ \emph {et~al.}(2022)\citenamefont {Xu},
  \citenamefont {Bao},\ and\ \citenamefont {Liew}}]{PhysRevB.106.L201302}%
  \BibitemOpen
  \bibfield  {author} {\bibinfo {author} {\bibfnamefont {X.}~\bibnamefont
  {Xu}}, \bibinfo {author} {\bibfnamefont {R.}~\bibnamefont {Bao}}, \ and\
  \bibinfo {author} {\bibfnamefont {T.~C.~H.}\ \bibnamefont {Liew}},\ }\href
  {\doibase 10.1103/PhysRevB.106.L201302} {\bibfield  {journal} {\bibinfo
  {journal} {Phys. Rev. B}\ }\textbf {\bibinfo {volume} {106}},\ \bibinfo
  {pages} {L201302} (\bibinfo {year} {2022})}\BibitemShut {NoStop}%
\bibitem [{\citenamefont {Comaron}\ \emph {et~al.}(2020)\citenamefont
  {Comaron}, \citenamefont {Shahnazaryan}, \citenamefont {Brzezicki},
  \citenamefont {Hyart},\ and\ \citenamefont
  {Matuszewski}}]{PhysRevResearch.2.022051}%
  \BibitemOpen
  \bibfield  {author} {\bibinfo {author} {\bibfnamefont {P.}~\bibnamefont
  {Comaron}}, \bibinfo {author} {\bibfnamefont {V.}~\bibnamefont
  {Shahnazaryan}}, \bibinfo {author} {\bibfnamefont {W.}~\bibnamefont
  {Brzezicki}}, \bibinfo {author} {\bibfnamefont {T.}~\bibnamefont {Hyart}}, \
  and\ \bibinfo {author} {\bibfnamefont {M.}~\bibnamefont {Matuszewski}},\
  }\href {\doibase 10.1103/PhysRevResearch.2.022051} {\bibfield  {journal}
  {\bibinfo  {journal} {Phys. Rev. Res.}\ }\textbf {\bibinfo {volume} {2}},\
  \bibinfo {pages} {022051} (\bibinfo {year} {2020})}\BibitemShut {NoStop}%
\bibitem [{\citenamefont {Yao}\ and\ \citenamefont
  {Wang}(2018)}]{PhysRevLett.121.086803}%
  \BibitemOpen
  \bibfield  {author} {\bibinfo {author} {\bibfnamefont {S.}~\bibnamefont
  {Yao}}\ and\ \bibinfo {author} {\bibfnamefont {Z.}~\bibnamefont {Wang}},\
  }\href {\doibase 10.1103/PhysRevLett.121.086803} {\bibfield  {journal}
  {\bibinfo  {journal} {Phys. Rev. Lett.}\ }\textbf {\bibinfo {volume} {121}},\
  \bibinfo {pages} {086803} (\bibinfo {year} {2018})}\BibitemShut {NoStop}%
\bibitem [{\citenamefont {Lee}\ and\ \citenamefont
  {Thomale}(2019)}]{PhysRevB.99.201103}%
  \BibitemOpen
  \bibfield  {author} {\bibinfo {author} {\bibfnamefont {C.~H.}\ \bibnamefont
  {Lee}}\ and\ \bibinfo {author} {\bibfnamefont {R.}~\bibnamefont {Thomale}},\
  }\href {\doibase 10.1103/PhysRevB.99.201103} {\bibfield  {journal} {\bibinfo
  {journal} {Phys. Rev. B}\ }\textbf {\bibinfo {volume} {99}},\ \bibinfo
  {pages} {201103} (\bibinfo {year} {2019})}\BibitemShut {NoStop}%
\bibitem [{\citenamefont {Yi}\ and\ \citenamefont
  {Yang}(2020)}]{PhysRevLett.125.186802}%
  \BibitemOpen
  \bibfield  {author} {\bibinfo {author} {\bibfnamefont {Y.}~\bibnamefont
  {Yi}}\ and\ \bibinfo {author} {\bibfnamefont {Z.}~\bibnamefont {Yang}},\
  }\href {\doibase 10.1103/PhysRevLett.125.186802} {\bibfield  {journal}
  {\bibinfo  {journal} {Phys. Rev. Lett.}\ }\textbf {\bibinfo {volume} {125}},\
  \bibinfo {pages} {186802} (\bibinfo {year} {2020})}\BibitemShut {NoStop}%
\bibitem [{\citenamefont {Li}\ \emph {et~al.}(2020)\citenamefont {Li},
  \citenamefont {Lee}, \citenamefont {Mu},\ and\ \citenamefont
  {Gong}}]{Li:2020vi}%
  \BibitemOpen
  \bibfield  {author} {\bibinfo {author} {\bibfnamefont {L.}~\bibnamefont
  {Li}}, \bibinfo {author} {\bibfnamefont {C.~H.}\ \bibnamefont {Lee}},
  \bibinfo {author} {\bibfnamefont {S.}~\bibnamefont {Mu}}, \ and\ \bibinfo
  {author} {\bibfnamefont {J.}~\bibnamefont {Gong}},\ }\href {\doibase
  10.1038/s41467-020-18917-4} {\bibfield  {journal} {\bibinfo  {journal}
  {Nature Communications}\ }\textbf {\bibinfo {volume} {11}},\ \bibinfo {pages}
  {5491} (\bibinfo {year} {2020})}\BibitemShut {NoStop}%
\bibitem [{\citenamefont {Zhu}\ \emph {et~al.}(2021)\citenamefont {Zhu},
  \citenamefont {Teo}, \citenamefont {Li},\ and\ \citenamefont
  {Gong}}]{PhysRevB.103.195414}%
  \BibitemOpen
  \bibfield  {author} {\bibinfo {author} {\bibfnamefont {W.}~\bibnamefont
  {Zhu}}, \bibinfo {author} {\bibfnamefont {W.~X.}\ \bibnamefont {Teo}},
  \bibinfo {author} {\bibfnamefont {L.}~\bibnamefont {Li}}, \ and\ \bibinfo
  {author} {\bibfnamefont {J.}~\bibnamefont {Gong}},\ }\href {\doibase
  10.1103/PhysRevB.103.195414} {\bibfield  {journal} {\bibinfo  {journal}
  {Phys. Rev. B}\ }\textbf {\bibinfo {volume} {103}},\ \bibinfo {pages}
  {195414} (\bibinfo {year} {2021})}\BibitemShut {NoStop}%
\bibitem [{\citenamefont {Wang}\ \emph {et~al.}(2022)\citenamefont {Wang},
  \citenamefont {Wang},\ and\ \citenamefont {Ma}}]{Wang:2022vc}%
  \BibitemOpen
  \bibfield  {author} {\bibinfo {author} {\bibfnamefont {W.}~\bibnamefont
  {Wang}}, \bibinfo {author} {\bibfnamefont {X.}~\bibnamefont {Wang}}, \ and\
  \bibinfo {author} {\bibfnamefont {G.}~\bibnamefont {Ma}},\ }\href {\doibase
  10.1038/s41586-022-04929-1} {\bibfield  {journal} {\bibinfo  {journal}
  {Nature}\ }\textbf {\bibinfo {volume} {608}},\ \bibinfo {pages} {50}
  (\bibinfo {year} {2022})}\BibitemShut {NoStop}%
\bibitem [{\citenamefont {Hohenberg}(1967)}]{PhysRev.158.383}%
  \BibitemOpen
  \bibfield  {author} {\bibinfo {author} {\bibfnamefont {P.~C.}\ \bibnamefont
  {Hohenberg}},\ }\href {\doibase 10.1103/PhysRev.158.383} {\bibfield
  {journal} {\bibinfo  {journal} {Phys. Rev.}\ }\textbf {\bibinfo {volume}
  {158}},\ \bibinfo {pages} {383} (\bibinfo {year} {1967})}\BibitemShut
  {NoStop}%
\bibitem [{\citenamefont {Fontaine}\ \emph {et~al.}(2022)\citenamefont
  {Fontaine}, \citenamefont {Squizzato}, \citenamefont {Baboux}, \citenamefont
  {Amelio}, \citenamefont {Lema{\^\i}tre}, \citenamefont {Morassi},
  \citenamefont {Sagnes}, \citenamefont {Le~Gratiet}, \citenamefont {Harouri},
  \citenamefont {Wouters}, \citenamefont {Carusotto}, \citenamefont {Amo},
  \citenamefont {Richard}, \citenamefont {Minguzzi}, \citenamefont {Canet},
  \citenamefont {Ravets},\ and\ \citenamefont {Bloch}}]{Fontaine:2022uk}%
  \BibitemOpen
  \bibfield  {author} {\bibinfo {author} {\bibfnamefont {Q.}~\bibnamefont
  {Fontaine}}, \bibinfo {author} {\bibfnamefont {D.}~\bibnamefont {Squizzato}},
  \bibinfo {author} {\bibfnamefont {F.}~\bibnamefont {Baboux}}, \bibinfo
  {author} {\bibfnamefont {I.}~\bibnamefont {Amelio}}, \bibinfo {author}
  {\bibfnamefont {A.}~\bibnamefont {Lema{\^\i}tre}}, \bibinfo {author}
  {\bibfnamefont {M.}~\bibnamefont {Morassi}}, \bibinfo {author} {\bibfnamefont
  {I.}~\bibnamefont {Sagnes}}, \bibinfo {author} {\bibfnamefont
  {L.}~\bibnamefont {Le~Gratiet}}, \bibinfo {author} {\bibfnamefont
  {A.}~\bibnamefont {Harouri}}, \bibinfo {author} {\bibfnamefont
  {M.}~\bibnamefont {Wouters}}, \bibinfo {author} {\bibfnamefont
  {I.}~\bibnamefont {Carusotto}}, \bibinfo {author} {\bibfnamefont
  {A.}~\bibnamefont {Amo}}, \bibinfo {author} {\bibfnamefont {M.}~\bibnamefont
  {Richard}}, \bibinfo {author} {\bibfnamefont {A.}~\bibnamefont {Minguzzi}},
  \bibinfo {author} {\bibfnamefont {L.}~\bibnamefont {Canet}}, \bibinfo
  {author} {\bibfnamefont {S.}~\bibnamefont {Ravets}}, \ and\ \bibinfo {author}
  {\bibfnamefont {J.}~\bibnamefont {Bloch}},\ }\href {\doibase
  10.1038/s41586-022-05001-8} {\bibfield  {journal} {\bibinfo  {journal}
  {Nature}\ }\textbf {\bibinfo {volume} {608}},\ \bibinfo {pages} {687}
  (\bibinfo {year} {2022})}\BibitemShut {NoStop}%
\bibitem [{\citenamefont {Kim}\ \emph {et~al.}(2016)\citenamefont {Kim},
  \citenamefont {Zhang}, \citenamefont {Wang}, \citenamefont {Fischer},
  \citenamefont {Brodbeck}, \citenamefont {Kamp}, \citenamefont {Schneider},
  \citenamefont {H\"ofling},\ and\ \citenamefont {Deng}}]{PhysRevX.6.011026}%
  \BibitemOpen
  \bibfield  {author} {\bibinfo {author} {\bibfnamefont {S.}~\bibnamefont
  {Kim}}, \bibinfo {author} {\bibfnamefont {B.}~\bibnamefont {Zhang}}, \bibinfo
  {author} {\bibfnamefont {Z.}~\bibnamefont {Wang}}, \bibinfo {author}
  {\bibfnamefont {J.}~\bibnamefont {Fischer}}, \bibinfo {author} {\bibfnamefont
  {S.}~\bibnamefont {Brodbeck}}, \bibinfo {author} {\bibfnamefont
  {M.}~\bibnamefont {Kamp}}, \bibinfo {author} {\bibfnamefont {C.}~\bibnamefont
  {Schneider}}, \bibinfo {author} {\bibfnamefont {S.}~\bibnamefont
  {H\"ofling}}, \ and\ \bibinfo {author} {\bibfnamefont {H.}~\bibnamefont
  {Deng}},\ }\href {\doibase 10.1103/PhysRevX.6.011026} {\bibfield  {journal}
  {\bibinfo  {journal} {Phys. Rev. X}\ }\textbf {\bibinfo {volume} {6}},\
  \bibinfo {pages} {011026} (\bibinfo {year} {2016})}\BibitemShut {NoStop}%
\bibitem [{\citenamefont {Wurdack}\ \emph {et~al.}(2022)\citenamefont
  {Wurdack}, \citenamefont {Estrecho}, \citenamefont {Todd}, \citenamefont
  {Schneider}, \citenamefont {Truscott},\ and\ \citenamefont
  {Ostrovskaya}}]{PhysRevLett.129.147402}%
  \BibitemOpen
  \bibfield  {author} {\bibinfo {author} {\bibfnamefont {M.}~\bibnamefont
  {Wurdack}}, \bibinfo {author} {\bibfnamefont {E.}~\bibnamefont {Estrecho}},
  \bibinfo {author} {\bibfnamefont {S.}~\bibnamefont {Todd}}, \bibinfo {author}
  {\bibfnamefont {C.}~\bibnamefont {Schneider}}, \bibinfo {author}
  {\bibfnamefont {A.~G.}\ \bibnamefont {Truscott}}, \ and\ \bibinfo {author}
  {\bibfnamefont {E.~A.}\ \bibnamefont {Ostrovskaya}},\ }\href {\doibase
  10.1103/PhysRevLett.129.147402} {\bibfield  {journal} {\bibinfo  {journal}
  {Phys. Rev. Lett.}\ }\textbf {\bibinfo {volume} {129}},\ \bibinfo {pages}
  {147402} (\bibinfo {year} {2022})}\BibitemShut {NoStop}%
\bibitem [{\citenamefont {Krizhanovskii}\ \emph {et~al.}(2009)\citenamefont
  {Krizhanovskii}, \citenamefont {Lagoudakis}, \citenamefont {Wouters},
  \citenamefont {Pietka}, \citenamefont {Bradley}, \citenamefont {Guda},
  \citenamefont {Whittaker}, \citenamefont {Skolnick}, \citenamefont
  {Deveaud-Pl\'edran}, \citenamefont {Richard}, \citenamefont {Andr\'e},\ and\
  \citenamefont {Dang}}]{PhysRevB.80.045317}%
  \BibitemOpen
  \bibfield  {author} {\bibinfo {author} {\bibfnamefont {D.~N.}\ \bibnamefont
  {Krizhanovskii}}, \bibinfo {author} {\bibfnamefont {K.~G.}\ \bibnamefont
  {Lagoudakis}}, \bibinfo {author} {\bibfnamefont {M.}~\bibnamefont {Wouters}},
  \bibinfo {author} {\bibfnamefont {B.}~\bibnamefont {Pietka}}, \bibinfo
  {author} {\bibfnamefont {R.~A.}\ \bibnamefont {Bradley}}, \bibinfo {author}
  {\bibfnamefont {K.}~\bibnamefont {Guda}}, \bibinfo {author} {\bibfnamefont
  {D.~M.}\ \bibnamefont {Whittaker}}, \bibinfo {author} {\bibfnamefont {M.~S.}\
  \bibnamefont {Skolnick}}, \bibinfo {author} {\bibfnamefont {B.}~\bibnamefont
  {Deveaud-Pl\'edran}}, \bibinfo {author} {\bibfnamefont {M.}~\bibnamefont
  {Richard}}, \bibinfo {author} {\bibfnamefont {R.}~\bibnamefont {Andr\'e}}, \
  and\ \bibinfo {author} {\bibfnamefont {L.~S.}\ \bibnamefont {Dang}},\ }\href
  {\doibase 10.1103/PhysRevB.80.045317} {\bibfield  {journal} {\bibinfo
  {journal} {Phys. Rev. B}\ }\textbf {\bibinfo {volume} {80}},\ \bibinfo
  {pages} {045317} (\bibinfo {year} {2009})}\BibitemShut {NoStop}%
\bibitem [{\citenamefont {Manni}\ \emph {et~al.}(2011)\citenamefont {Manni},
  \citenamefont {Lagoudakis}, \citenamefont {Pietka}, \citenamefont
  {Fontanesi}, \citenamefont {Wouters}, \citenamefont {Savona}, \citenamefont
  {Andr\'e},\ and\ \citenamefont {Deveaud-Pl\'edran}}]{PhysRevLett.106.176401}%
  \BibitemOpen
  \bibfield  {author} {\bibinfo {author} {\bibfnamefont {F.}~\bibnamefont
  {Manni}}, \bibinfo {author} {\bibfnamefont {K.~G.}\ \bibnamefont
  {Lagoudakis}}, \bibinfo {author} {\bibfnamefont {B.}~\bibnamefont {Pietka}},
  \bibinfo {author} {\bibfnamefont {L.}~\bibnamefont {Fontanesi}}, \bibinfo
  {author} {\bibfnamefont {M.}~\bibnamefont {Wouters}}, \bibinfo {author}
  {\bibfnamefont {V.}~\bibnamefont {Savona}}, \bibinfo {author} {\bibfnamefont
  {R.}~\bibnamefont {Andr\'e}}, \ and\ \bibinfo {author} {\bibfnamefont
  {B.}~\bibnamefont {Deveaud-Pl\'edran}},\ }\href {\doibase
  10.1103/PhysRevLett.106.176401} {\bibfield  {journal} {\bibinfo  {journal}
  {Phys. Rev. Lett.}\ }\textbf {\bibinfo {volume} {106}},\ \bibinfo {pages}
  {176401} (\bibinfo {year} {2011})}\BibitemShut {NoStop}%
\bibitem [{\citenamefont {T\"{o}pfer}\ \emph {et~al.}(2021)\citenamefont
  {T\"{o}pfer}, \citenamefont {Chatzopoulos}, \citenamefont {Sigurdsson},
  \citenamefont {Cookson}, \citenamefont {Rubo},\ and\ \citenamefont
  {Lagoudakis}}]{Topfer:21}%
  \BibitemOpen
  \bibfield  {author} {\bibinfo {author} {\bibfnamefont {J.~D.}\ \bibnamefont
  {T\"{o}pfer}}, \bibinfo {author} {\bibfnamefont {I.}~\bibnamefont
  {Chatzopoulos}}, \bibinfo {author} {\bibfnamefont {H.}~\bibnamefont
  {Sigurdsson}}, \bibinfo {author} {\bibfnamefont {T.}~\bibnamefont {Cookson}},
  \bibinfo {author} {\bibfnamefont {Y.~G.}\ \bibnamefont {Rubo}}, \ and\
  \bibinfo {author} {\bibfnamefont {P.~G.}\ \bibnamefont {Lagoudakis}},\ }\href
  {\doibase 10.1364/OPTICA.409976} {\bibfield  {journal} {\bibinfo  {journal}
  {Optica}\ }\textbf {\bibinfo {volume} {8}},\ \bibinfo {pages} {106} (\bibinfo
  {year} {2021})}\BibitemShut {NoStop}%
\bibitem [{\citenamefont {Kr\'ol}\ \emph {et~al.}(2021)\citenamefont {Kr\'ol},
  \citenamefont {Rechci\ifmmode~\acute{n}\else \'{n}\fi{}ska}, \citenamefont
  {Sigurdsson}, \citenamefont {Oliwa}, \citenamefont {Mazur}, \citenamefont
  {Morawiak}, \citenamefont {Piecek}, \citenamefont {Kula}, \citenamefont
  {Lagoudakis}, \citenamefont {Matuszewski}, \citenamefont {Bardyszewski},
  \citenamefont {Pi\ifmmode~\mbox{\k{e}}\else \k{e}\fi{}tka},\ and\
  \citenamefont {Szczytko}}]{PhysRevLett.127.190401}%
  \BibitemOpen
  \bibfield  {author} {\bibinfo {author} {\bibfnamefont {M.}~\bibnamefont
  {Kr\'ol}}, \bibinfo {author} {\bibfnamefont {K.}~\bibnamefont
  {Rechci\ifmmode~\acute{n}\else \'{n}\fi{}ska}}, \bibinfo {author}
  {\bibfnamefont {H.}~\bibnamefont {Sigurdsson}}, \bibinfo {author}
  {\bibfnamefont {P.}~\bibnamefont {Oliwa}}, \bibinfo {author} {\bibfnamefont
  {R.}~\bibnamefont {Mazur}}, \bibinfo {author} {\bibfnamefont
  {P.}~\bibnamefont {Morawiak}}, \bibinfo {author} {\bibfnamefont
  {W.}~\bibnamefont {Piecek}}, \bibinfo {author} {\bibfnamefont
  {P.}~\bibnamefont {Kula}}, \bibinfo {author} {\bibfnamefont {P.~G.}\
  \bibnamefont {Lagoudakis}}, \bibinfo {author} {\bibfnamefont
  {M.}~\bibnamefont {Matuszewski}}, \bibinfo {author} {\bibfnamefont
  {W.}~\bibnamefont {Bardyszewski}}, \bibinfo {author} {\bibfnamefont
  {B.}~\bibnamefont {Pi\ifmmode~\mbox{\k{e}}\else \k{e}\fi{}tka}}, \ and\
  \bibinfo {author} {\bibfnamefont {J.}~\bibnamefont {Szczytko}},\ }\href
  {\doibase 10.1103/PhysRevLett.127.190401} {\bibfield  {journal} {\bibinfo
  {journal} {Phys. Rev. Lett.}\ }\textbf {\bibinfo {volume} {127}},\ \bibinfo
  {pages} {190401} (\bibinfo {year} {2021})}\BibitemShut {NoStop}%
\bibitem [{\citenamefont {Li}\ \emph {et~al.}(2022)\citenamefont {Li},
  \citenamefont {Ma}, \citenamefont {Zhai}, \citenamefont {Gao}, \citenamefont
  {Dai}, \citenamefont {Schumacher},\ and\ \citenamefont {Gao}}]{Li:2022uu}%
  \BibitemOpen
  \bibfield  {author} {\bibinfo {author} {\bibfnamefont {Y.}~\bibnamefont
  {Li}}, \bibinfo {author} {\bibfnamefont {X.}~\bibnamefont {Ma}}, \bibinfo
  {author} {\bibfnamefont {X.}~\bibnamefont {Zhai}}, \bibinfo {author}
  {\bibfnamefont {M.}~\bibnamefont {Gao}}, \bibinfo {author} {\bibfnamefont
  {H.}~\bibnamefont {Dai}}, \bibinfo {author} {\bibfnamefont {S.}~\bibnamefont
  {Schumacher}}, \ and\ \bibinfo {author} {\bibfnamefont {T.}~\bibnamefont
  {Gao}},\ }\href {\doibase 10.1038/s41467-022-31529-4} {\bibfield  {journal}
  {\bibinfo  {journal} {Nature Communications}\ }\textbf {\bibinfo {volume}
  {13}},\ \bibinfo {pages} {3785} (\bibinfo {year} {2022})}\BibitemShut
  {NoStop}%
\bibitem [{\citenamefont {Kokhanchik}\ \emph {et~al.}(2023)\citenamefont
  {Kokhanchik}, \citenamefont {Solnyshkov},\ and\ \citenamefont
  {Malpuech}}]{kokhanchik2023nonhermitian}%
  \BibitemOpen
  \bibfield  {author} {\bibinfo {author} {\bibfnamefont {P.}~\bibnamefont
  {Kokhanchik}}, \bibinfo {author} {\bibfnamefont {D.}~\bibnamefont
  {Solnyshkov}}, \ and\ \bibinfo {author} {\bibfnamefont {G.}~\bibnamefont
  {Malpuech}},\ }\href@noop {} {\bibfield  {journal} {\bibinfo  {journal}
  {arxiv}\ }\textbf {\bibinfo {volume} {2303.08483}} (\bibinfo {year}
  {2023})}\BibitemShut {NoStop}%
\bibitem [{\citenamefont {Fischer}\ \emph {et~al.}(2014)\citenamefont
  {Fischer}, \citenamefont {Savenko}, \citenamefont {Fraser}, \citenamefont
  {Holzinger}, \citenamefont {Brodbeck}, \citenamefont {Kamp}, \citenamefont
  {Shelykh}, \citenamefont {Schneider},\ and\ \citenamefont
  {H\"ofling}}]{PhysRevLett.113.203902}%
  \BibitemOpen
  \bibfield  {author} {\bibinfo {author} {\bibfnamefont {J.}~\bibnamefont
  {Fischer}}, \bibinfo {author} {\bibfnamefont {I.~G.}\ \bibnamefont
  {Savenko}}, \bibinfo {author} {\bibfnamefont {M.~D.}\ \bibnamefont {Fraser}},
  \bibinfo {author} {\bibfnamefont {S.}~\bibnamefont {Holzinger}}, \bibinfo
  {author} {\bibfnamefont {S.}~\bibnamefont {Brodbeck}}, \bibinfo {author}
  {\bibfnamefont {M.}~\bibnamefont {Kamp}}, \bibinfo {author} {\bibfnamefont
  {I.~A.}\ \bibnamefont {Shelykh}}, \bibinfo {author} {\bibfnamefont
  {C.}~\bibnamefont {Schneider}}, \ and\ \bibinfo {author} {\bibfnamefont
  {S.}~\bibnamefont {H\"ofling}},\ }\href {\doibase
  10.1103/PhysRevLett.113.203902} {\bibfield  {journal} {\bibinfo  {journal}
  {Phys. Rev. Lett.}\ }\textbf {\bibinfo {volume} {113}},\ \bibinfo {pages}
  {203902} (\bibinfo {year} {2014})}\BibitemShut {NoStop}%
\bibitem [{\citenamefont {Gong}\ \emph {et~al.}(2018)\citenamefont {Gong},
  \citenamefont {Ashida}, \citenamefont {Kawabata}, \citenamefont {Takasan},
  \citenamefont {Higashikawa},\ and\ \citenamefont {Ueda}}]{PhysRevX.8.031079}%
  \BibitemOpen
  \bibfield  {author} {\bibinfo {author} {\bibfnamefont {Z.}~\bibnamefont
  {Gong}}, \bibinfo {author} {\bibfnamefont {Y.}~\bibnamefont {Ashida}},
  \bibinfo {author} {\bibfnamefont {K.}~\bibnamefont {Kawabata}}, \bibinfo
  {author} {\bibfnamefont {K.}~\bibnamefont {Takasan}}, \bibinfo {author}
  {\bibfnamefont {S.}~\bibnamefont {Higashikawa}}, \ and\ \bibinfo {author}
  {\bibfnamefont {M.}~\bibnamefont {Ueda}},\ }\href {\doibase
  10.1103/PhysRevX.8.031079} {\bibfield  {journal} {\bibinfo  {journal} {Phys.
  Rev. X}\ }\textbf {\bibinfo {volume} {8}},\ \bibinfo {pages} {031079}
  (\bibinfo {year} {2018})}\BibitemShut {NoStop}%
\bibitem [{\citenamefont {Kartashov}\ and\ \citenamefont
  {Skryabin}(2019)}]{PhysRevLett.122.083902}%
  \BibitemOpen
  \bibfield  {author} {\bibinfo {author} {\bibfnamefont {Y.~V.}\ \bibnamefont
  {Kartashov}}\ and\ \bibinfo {author} {\bibfnamefont {D.~V.}\ \bibnamefont
  {Skryabin}},\ }\href {\doibase 10.1103/PhysRevLett.122.083902} {\bibfield
  {journal} {\bibinfo  {journal} {Phys. Rev. Lett.}\ }\textbf {\bibinfo
  {volume} {122}},\ \bibinfo {pages} {083902} (\bibinfo {year}
  {2019})}\BibitemShut {NoStop}%
\bibitem [{\citenamefont {St-Jean}\ \emph {et~al.}(2017)\citenamefont
  {St-Jean}, \citenamefont {Goblot}, \citenamefont {Galopin}, \citenamefont
  {Lema{\^\i}tre}, \citenamefont {Ozawa}, \citenamefont {Le~Gratiet},
  \citenamefont {Sagnes}, \citenamefont {Bloch},\ and\ \citenamefont
  {Amo}}]{St-Jean:2017ws}%
  \BibitemOpen
  \bibfield  {author} {\bibinfo {author} {\bibfnamefont {P.}~\bibnamefont
  {St-Jean}}, \bibinfo {author} {\bibfnamefont {V.}~\bibnamefont {Goblot}},
  \bibinfo {author} {\bibfnamefont {E.}~\bibnamefont {Galopin}}, \bibinfo
  {author} {\bibfnamefont {A.}~\bibnamefont {Lema{\^\i}tre}}, \bibinfo {author}
  {\bibfnamefont {T.}~\bibnamefont {Ozawa}}, \bibinfo {author} {\bibfnamefont
  {L.}~\bibnamefont {Le~Gratiet}}, \bibinfo {author} {\bibfnamefont
  {I.}~\bibnamefont {Sagnes}}, \bibinfo {author} {\bibfnamefont
  {J.}~\bibnamefont {Bloch}}, \ and\ \bibinfo {author} {\bibfnamefont
  {A.}~\bibnamefont {Amo}},\ }\href {\doibase 10.1038/s41566-017-0006-2}
  {\bibfield  {journal} {\bibinfo  {journal} {Nature Photonics}\ }\textbf
  {\bibinfo {volume} {11}},\ \bibinfo {pages} {651} (\bibinfo {year}
  {2017})}\BibitemShut {NoStop}%
\bibitem [{\citenamefont {Estrecho}\ \emph {et~al.}(2018)\citenamefont
  {Estrecho}, \citenamefont {Gao}, \citenamefont {Bobrovska}, \citenamefont
  {Fraser}, \citenamefont {Steger}, \citenamefont {Pfeiffer}, \citenamefont
  {West}, \citenamefont {Liew}, \citenamefont {Matuszewski}, \citenamefont
  {Snoke}, \citenamefont {Truscott},\ and\ \citenamefont
  {Ostrovskaya}}]{Estrecho:2018vv}%
  \BibitemOpen
  \bibfield  {author} {\bibinfo {author} {\bibfnamefont {E.}~\bibnamefont
  {Estrecho}}, \bibinfo {author} {\bibfnamefont {T.}~\bibnamefont {Gao}},
  \bibinfo {author} {\bibfnamefont {N.}~\bibnamefont {Bobrovska}}, \bibinfo
  {author} {\bibfnamefont {M.~D.}\ \bibnamefont {Fraser}}, \bibinfo {author}
  {\bibfnamefont {M.}~\bibnamefont {Steger}}, \bibinfo {author} {\bibfnamefont
  {L.}~\bibnamefont {Pfeiffer}}, \bibinfo {author} {\bibfnamefont
  {K.}~\bibnamefont {West}}, \bibinfo {author} {\bibfnamefont {T.~C.~H.}\
  \bibnamefont {Liew}}, \bibinfo {author} {\bibfnamefont {M.}~\bibnamefont
  {Matuszewski}}, \bibinfo {author} {\bibfnamefont {D.~W.}\ \bibnamefont
  {Snoke}}, \bibinfo {author} {\bibfnamefont {A.~G.}\ \bibnamefont {Truscott}},
  \ and\ \bibinfo {author} {\bibfnamefont {E.~A.}\ \bibnamefont
  {Ostrovskaya}},\ }\href {\doibase 10.1038/s41467-018-05349-4} {\bibfield
  {journal} {\bibinfo  {journal} {Nature Communications}\ }\textbf {\bibinfo
  {volume} {9}},\ \bibinfo {pages} {2944} (\bibinfo {year} {2018})}\BibitemShut
  {NoStop}%
\bibitem [{\citenamefont {Tsyplyatyev}\ and\ \citenamefont
  {Whittaker}(2012)}]{https://doi.org/10.1002/pssb.201248074}%
  \BibitemOpen
  \bibfield  {author} {\bibinfo {author} {\bibfnamefont {O.}~\bibnamefont
  {Tsyplyatyev}}\ and\ \bibinfo {author} {\bibfnamefont {D.~M.}\ \bibnamefont
  {Whittaker}},\ }\href {\doibase https://doi.org/10.1002/pssb.201248074}
  {\bibfield  {journal} {\bibinfo  {journal} {physica status solidi (b)}\
  }\textbf {\bibinfo {volume} {249}},\ \bibinfo {pages} {1692} (\bibinfo {year}
  {2012})}\BibitemShut {NoStop}%
\end{thebibliography}%
\bibliographystyle{apsrev4-1}

\end{document}